\documentclass[aps, superscriptaddress,  twocolumn]{revtex4}
\usepackage{CJK}
\usepackage{mathtools}
\usepackage[normalem]{ulem}
\usepackage{graphicx}
\usepackage{dcolumn}
\usepackage[mathlines]{lineno}
\usepackage{hyperref}
\usepackage{bm}
\usepackage{amssymb}
\usepackage{amsmath}
\usepackage{siunitx}
\usepackage{braket}
\usepackage{color}
\usepackage{xcolor}
\usepackage{soul} 
\usepackage{graphics}
\usepackage{bbm}

\begin{document}
\title{Tunable spatio-spectral Target Skyrmions and topological multiplexing}

\author{Pedro Ornelas}
\affiliation{School of Physics, University of the Witwatersrand, Private Bag 3, Wits 2050, South Africa}

\author{Niladri Modak}
\affiliation{Tampere University, Photonics Laboratory, Physics Unit, Tampere, FI-33720, Finland}

\author{Oussama Korichi}
\affiliation{Tampere University, Photonics Laboratory, Physics Unit, Tampere, FI-33720, Finland}

\author{Isaac Nape}
\affiliation{School of Physics, University of the Witwatersrand, Private Bag 3, Wits 2050, South Africa}

\author{Andrew Forbes}
\affiliation{School of Physics, University of the Witwatersrand, Private Bag 3, Wits 2050, South Africa}

\author{Robert Fickler}
\email{robert.fickler@tuni.fi}
\affiliation{Tampere University, Photonics Laboratory, Physics Unit, Tampere, FI-33720, Finland}

\begin{abstract}
\noindent \textbf{Optical Skyrmions have recently garnered much interest providing a potential avenue for high capacity, robust topological information transfer. Typically, Skyrmions are derived from the coupling of just two degrees of freedom (DoFs) limiting their versatility. In this work we realize spatio-spectral Skyrmions derived from the non-separability between three DoFs: wavelength, space and polarization. A compact and simple technique is used to generate the spatio-spectral vector beams (SSVB) carrying the desired Skyrmionic structure,  offering simple pathways for complex Skyrmionic beam design. The topological structure, witnessed through a map between the spatio-spectral plane and the Poincar\'e sphere, exhibits an additional tunable $k\pi$ parameter thereby enhancing the number of controllable DoFs. Our three DoF construction allows us to propose a novel topological multiplexing strategy that independently encodes different Skyrmion numbers at different radii of the field. We experimentally demonstrate the practicality of this approach by transmitting and receiving three distinct Skyrmion numbers encoded into a single topological field, for the first form of mode division multiplexing with Skyrmion topology. This work opens up new avenues for dense information encoding using multiple topological channels encoded in a single light field.}
\end{abstract}

\maketitle



\section{Introduction}

\noindent Skyrmions are topologically stable field configurations characterized by a topological invariant, the Skyrmion number, defined as the number of times its domain wraps a selected target space. Whilst initially postulated by Tony Skyrme in the early 1960s as a way to describe sub-atomic particles as excitations of a basic pion field theory \cite{skyrme1962unified,ZAHED19861, Naya2018Skyrmions,eisenberg1981nucleon}, the generality of their definition has since allowed for their emergence within diverse physical systems including Bose-Einstein condensates \cite{leslie2009creation}, chiral liquid crystals \cite{ackerman2015self}, acoustic waves \cite{ge2021observation, muelas2022observation}, and water-waves \cite{wang2025topological}. In particular the past decade has witnessed remarkable advancements in the generation and control of optical Skyrmions across diverse photonic systems having been generated within evanescent waves \cite{tsesses2018optical}, 
non-paraxial light \cite{lei2025skyrmionic, du2019deep}, in chiral liquid-crystal-filled micro-cavities \cite{krol2021observation}, and as toroidal pulses \cite{shen2021supertoroidal}. Notably, free-space Stokes skyrmions generated as coherent, paraxial beams \cite{cisowski2023building, shen2022generation, gao2020paraxial, singh2023synthetic, teng2023physical, shen2021topological}, single photon spin-orbit coupled states \cite{ma2025nanophotonic, koni2025dual} and derived from correlations between distant entangled photons \cite{ornelas2024non} 
have risen as prominent candidates for information encoding schemes offering novel pathways for high capacity \cite{wang2025generation, zeng2025tailoring} optical communication and optical computing \cite{wang2025perturbation} with demonstrated resilience through diverse noisy environments \cite{wang2024topological,  ornelas2025topological, de2025quantum, wang2025topological, guo2026topological}. \\

\noindent Development of sophisticated tools to control and manipulate light's many DoFs has unlocked the ability to tailor high dimensional topological structures such as Skyrmionic Hopfions \cite{sugic2021particle} in three spatial dimensions, manipulate topological features within different fields of an optical system \cite{yao2024multi}, and most recently embed topologies between multiple DoFs simultaneously, achieved following the discovery of spatio-temporal optical vortices (STOVs) \cite{chong2020generation} enabling the generation of spatio-temportal optical Skyrmions \cite{teng2025construction}. However, whilst impressive these structures typically require complicated generation, detection and analysis techniques without fully exploiting their additional complexity for any additional utility. \\

\noindent Here we realize a novel optical Skyrmion derived from the non-separability between three DoFs - wavelength, space and polarization - which we term spatio-spectral target Skyrmions (SSTS). 
We create the desired Skyrmionic structure by encoding appropriate correlations into a spatio-spectral vector beam (SSVB) generated using a compact and simple generation technique 
offering an accessible and practical avenue for complex Skyrmionic beam generation. 
The topological structure of our generated SSVBs is revealed using a compact representation of the tripartite-like correlations embedded within the state, through a map between the spatio-spectral plane and the Poincar\'e sphere. 
Furthermore, by altering the periodicity of polarization oscillation within the spectral domain, we demonstrate the ability to control the number of full $k\pi$ rotations exhibited by the Skyrmionic mapping thereby offering an additional tunable mechanism. 
Lastly, we propose and demonstrate a novel topological multiplexing strategy which fully exploits all dimensions of our SSVB through independently encoding different Skyrmion numbers at different radii of the field. 
We show the practicality of this approach by realizing an SSVB characterized by three distinct Skyrmion numbers. Therefore, this work has potential to open up new avenues for dense information encoding using multiple topological channels encoded in a single light field. 
 

\begin{figure*}[t!]
\includegraphics[width=0.95\linewidth]{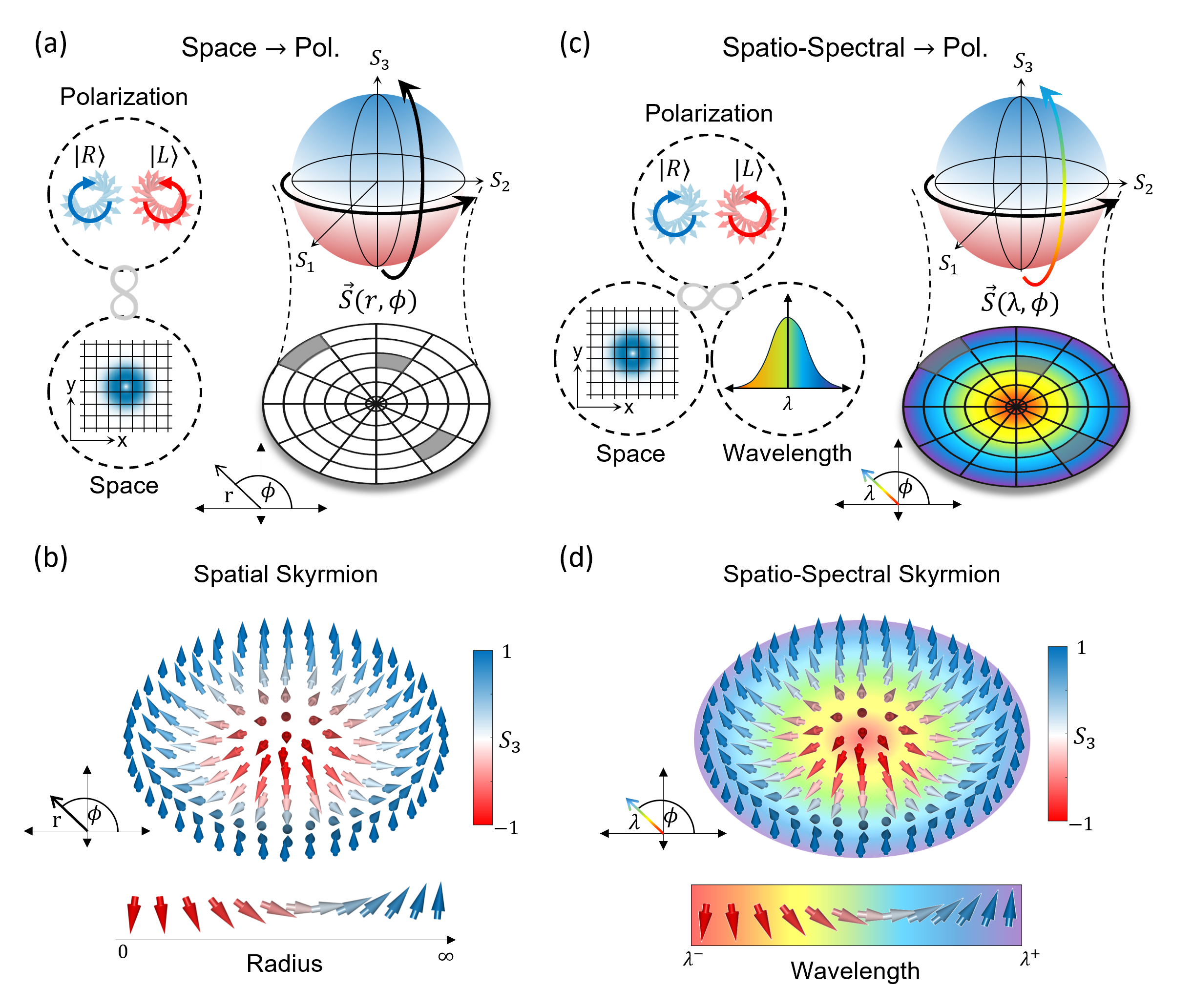}
\caption{\textbf{Concept.} 
(a) Typical vector beams carry correlations between space and polarization which can be expressed as a map, $\vec{S}(\vec{r})$, between the spatial plane, $R^2$, with coordinates $\vec{r}=(r,\phi)$ and the Poincar\'e sphere with coordinates $(S_1,S_2,S_3)$. (b) Representation of the vector field, $\vec{S}(\vec{r})$, of a typical Ne\'el type Skyrmion defined within the spatial domain where the polarization state rotates from the south to the north pole of the Poincar\'e sphere with a change in radius. 
(c) Spatio-spectral vector beams correlate space, wavelength and polarization yielding complex mapping structures. Replacing the radial coordinate, $r$, with a wavelength coordinate, $\lambda$, achieves the map, $S(\lambda, \phi)$, which maps the spatio-spectral plane to the Poincar\'e sphere. (d) Representation of the vector field, $\vec{S}(\lambda,\phi)$, of a typical Ne\'el type Skyrmion defined within the spatio-spectral domain where the polarization state rotates from the south to the north pole of the Poincar\'e sphere with a change in wavelength.}
\label{fig:Concept}
\end{figure*}

\section{Concept}

\noindent Figure 1. illustrates our framework for realizing optical Skyrmions with extended physical degrees of freedom by judiciously structuring the spatial, spectral and polarization components of light. This approach highlights how continuous degrees of freedom can be harnessed to engineer skyrmionic textures in which the spectral component of light now forms part of the physical mapping function that encodes the underlying topology.

A useful starting point is to consider the typical optical Skyrmionic constructions that are encoded within spin-textured fields. Such fields can be identified with a map, $\vec{S}(\vec{r}):R^2 \xrightarrow{N} S^2$, characterized by a Skyrmion number $N \in \mathbb{Z}$  that captures the underlying topology of the field. Here, $N$ measures the number of times the spatial plane, $R^2$, wraps the parameter sphere, $S^2$, connected by the vector field $\vec{S}(\vec{r})$. A popular example is that of a vector beam possessing a spatially-varying polarization profile where the embedded correlations between space and polarization are then typically captured by the normalized Stokes (equivalently spin) vector field, $\vec{S}(\vec{r})=(S_1(\vec{r}),S_2(\vec{r}),S_3(\vec{r}))$, which forms a map between the spatial plane with cylindrical coordinates $\vec{r}=(r,\phi)$ 
and the Poincar\'e sphere with coordinates $(S_1,S_2,S_3)$ as shown in Figure~\ref{fig:Concept} (a). A common example of such a map is shown in Figure~\ref{fig:Concept} (b) as a vector field with a Ne\'el type texture. Here the vector field rotates about the $S_3$ axis with changing azimuthal angle and it rotates from the south to the north pole with changing radius.\\

Under this framework, the introduction of wavelength as an additional correlated degree of freedom (DoF) is captured by the extension of the Stokes vector to $\vec{S}(\lambda, \vec{r})$. In this way, our mapping function for the topology is endowed with a wavelength/spectral coordinate, marking a clear departure from conventional skyrmion descriptions based purely on space-spin mappings. To this end, we combine all three DoFs, wavelength, space and polarization simultaneously to achieve an equivalent map to that of the typical (spatial) optical Skyrmion shown in Figure~\ref{fig:Concept} (a). Here the typical radial coordinate has been replaced by a wavelength coordinate, thus the new map is represented by the Stokes vector, $\vec{S}(\lambda, \phi)$ and forms a map between the spatio-spectral plane and the Poincar\'e sphere. Represented as a vector field, we observe now that whilst the azimuthal coordinate is still responsible for the orientation of the Stokes vector with respect to the $S_3$ axis, the vector rotates from the south to north pole with a change in wavelength from $\lambda^-$ to $\lambda^+$ as shown in Figure~\ref{fig:Concept} (c).
The resulting Neél-type Skrymion becomes visible for the vector field plotted against angle and wavelength as shown in Figure~\ref{fig:Concept} (d).\\



\begin{figure*}[t!]
\includegraphics[width=0.9\linewidth]{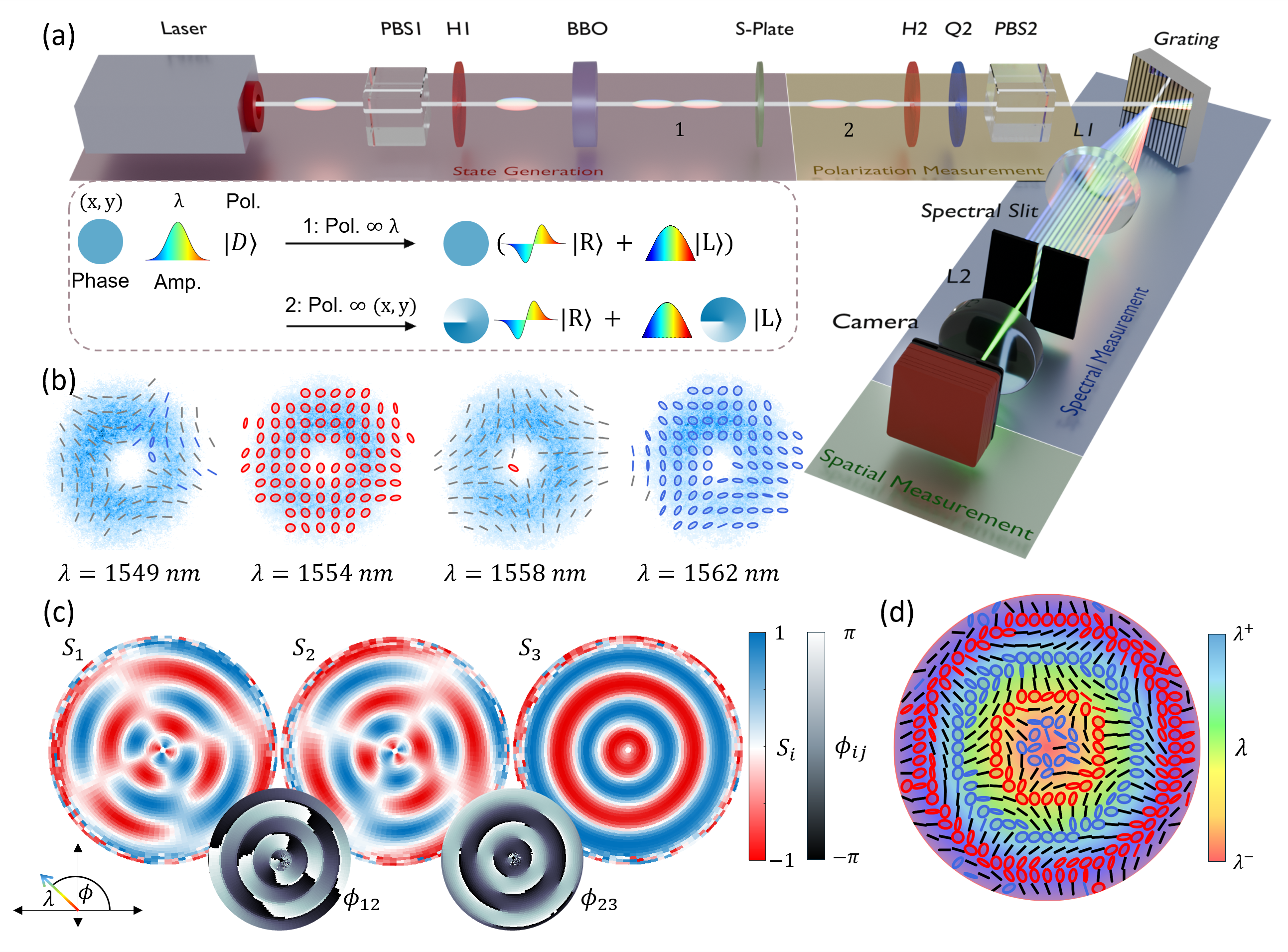}
\caption{\textbf{Experimental realization of SSVBs} (a) Schematic of the experimental setup for the generation and detection of spatio-spectral vector beams. The fs laser pulse polarization is set diagonal to the birefringent Beta-Barium Borate (BBO) crystal by the polarizing beam splitter (PBS1) and half-waveplate (H1). The BBO and a vortex half-wave retarder (S-Plate) imprint the desired polarization pattern in the spectral ($\lambda$) and spatial azimuthal ($\phi$) degrees of freedom. The SSVB is characterized by subsequent measurements of polarization with  H2, a quarter waveplate (Q2), and PBS2; spectrum via a home-built monochromator (grating, lenses L1, L2, and a spectral slit), and spatial profile with a camera. 
(b) The recorded spatial-polarization structures of the beam is presented for different wavelengths $\lambda\in\{1549,1554,1558,1562\}\ \text{nm}$. (c) The derived spatio-spectral Stokes vector elements $S_1$, $S_2$, and $S_3$ and associated Stokes phases $\Phi_{ij}=\arctan\left(\frac{S_j(\lambda,\phi)}{S_i(\lambda,\phi)}\right)$. (d) Polarization ellipses, derived from the Stokes vector elements, emphasizing the polarization and spatio-spectral correlations. The encoded colour is used to differentiate between linear (black), right (red) and left circular polarization (blue) states.}
\label{fig:Experiment}
\end{figure*}

\section{Results}

\noindent \textbf{Experiment.} To achieve the goal of demonstrating our proposed Skyrmionic topological states, we synthesize beams with nonseparable spatial, spectral and polarization components. In this work, the underlying fields are prepared through constructing a SSVB of the form

\begin{equation}
|\Psi\rangle = F_R(\lambda) G_R(\vec{r})|R\rangle + F_L(\lambda) G_L(\vec{r})|L\rangle 
\label{eq:SpatSpecVB}
\end{equation}

where $F_{R,L}(\lambda)$ and $G_{R,L}(\vec{r})$ are spectral and spatial functions, respectively, encoding the desired spatio-spectral correlations. To generate these tunable SSVBs we utilize an experiment inspired by the approach reported in \cite{kopf2023correlating, fickler2024higher} as shown in  Fig.~\ref{fig:Experiment} (a). In our experiment, we use Fourier limited laser pulses of wavelength $1560$ nm, and $220$ fs pulse duration. The pulse is then polarized diagonally using a polarizing beam splitter and half-wave plate 1 rotated at $22.5^{\circ}$ with respect to the axis of a $4$ mm thick birefringent material comprised of two Beta-Barium Borate (BBO) crystals with cut angles $41.8^\circ$ and $23.4^\circ$. This dual crystal configuration was used to minimize spatial walk-off by orientating the crystals $180^{\circ}$ with respect to one another. The BBO configuration coherently splits the pulse into two orthogonally polarized trailing pulses separated by $0.48$ ps. This delay in time results in a linearly varying phase between the two orthogonal polarizations across the wavelength spectrum, thus generating a spectral vector beam (SVB) exhibiting correlations between polarization and wavelength (See supplementary information for SVB characterization). 
Lastly, to generate a full SSVB of the form given in Eq.~\ref{eq:SpatSpecVB}, we pass the SVB through a zero-order vortex half-wave retarder (S-Plate) which encodes the desired spatial-polarization correlations. \\
To characterize our state, we perform simultaneous polarization, spectral and spatial measurements as depicted in Fig.~\ref{fig:Experiment} (a). The polarization measurement is performed using a second half-wave plate and quarter wave plate, with a polarizing-beam splitter which projects onto the horizontal polarization state. The spectral measurement is performed using a grating, a pair of lenses and a spectral slit. The grating and lens combination maps the wavelength to distinct positions at the focal plane of the first lens where a slit is placed to perform the desired spectral filtering. Subsequently, a camera placed at the focal plane of the second $L_2$ performs the spatial measurement.\\ 

This setup allows us to embed and measure non-trivial topological structures within the spatio-spectral domain. In our case, the temporal delay and spatially-varying phase retardance is introduced by the BBO crystal and S-Plate, respectively. The former results in a relative amplitude variation between the right and left circular components of the field across the spectral domain. The latter leads to a phase change between the circular polarizations along azimuthal angle of the field's transverse plane. The resultant SSVB possesses tripartite-like correlations which can be described through a surjective map between the spatio-spectral domain and the Poincar\'e sphere. The transformation that the state undergoes can be summarized as follows (we have omitted the Gaussian spectral, $F(\lambda)$, and spatial $G(r)$ envelope functions for simplicity)



\begin{eqnarray}
|D\rangle
&\xrightarrow{\text{BBO }}& \left(\cos(f(\lambda))|R\rangle + \sin(f(\lambda))|L\rangle\right)  \\
&\xrightarrow{\text{S-plate}}&
e^{ig(\phi)}\cos(f(\lambda))|R\rangle + \nonumber \\ 
&&e^{-ig(\phi)}\sin(f(\lambda))|L\rangle
\label{eq:FieldTransform1}
\end{eqnarray}
where $g(\phi) = 2q\phi$, $f(\lambda)=\omega \Delta t = \frac{2\pi c}{\lambda}\Delta t$ with $c$ being the speed of light in vacuum and $\Delta t$ being the delay imparted by the BBO crystal (more details provided in the supplementary material). The results shown in Fig.~\ref{fig:Experiment} (b) depict polarization ellipse textures overlaid on the total field intensity for wavelength projections, $\lambda=\{1549,1554, 1558, 1562 \}\text{nm}$. Here the observed polarization distribution oscillates between being homogeneously circularly polarized as shown for spectral projections $\lambda=\{1554,1562\}$nm and inhomogeneously linearly polarized as shown for spectral projections $\lambda=\{1549, 1562\}$nm). This is consistent with the expected behaviour of the field given in Eq.~\ref{eq:FieldTransform1} with changing wavelength. The field oscillates between two separable, circularly polarized scalar states, $\{|R\rangle,|L\rangle\}$ with maximally non-separable linearly polarized states of the form $\{e^{i\phi}|R\rangle + e^{-i\phi}|L\rangle\}$, found at intermediary wavelengths. The experimental Stokes vector, $\vec{S}(\lambda, r, \phi)$, can then be calculated by extracting its components given by $S_i(\lambda, r,\phi) = \langle\Psi|\sigma_i|\Psi\rangle$ where $i\in\{x,y,z\}$ and $\sigma_i$ are the typical Pauli-spin matrices. The theoretical, locally normalized $\left(\vec{S}\cdot\vec{S}=1\right)$ Stokes vector for the SSVB defined in Eq.~\ref{eq:FieldTransform1} is given by 
\begin{equation}
\vec{S}(\lambda,r,\phi) = \begin{bmatrix}\cos(2g(\phi)) \sin(2f(\lambda)) \\ 
\sin(2g(\phi)) \sin(2f(\lambda)) \\  \cos(2f(\lambda))\end{bmatrix}
\label{eq:SpatSpecStokesFP}
\end{equation}
Since the normalized Stokes vector exhibits no radial dependence, the Stokes vector can be collapsed into the spatio-spectral Stokes vector, $\vec{S}(\lambda,\phi)$, which involves replacing the radial coordinate by a wavelength coordinate, $\lambda \in [\lambda^-,\lambda^+]$, where $\lambda^-$ and  $\lambda^+$ form the lower and upper spectral bounds of our field. This scheme allows for the compact visualization of the spatio-spectral-polarization correlations of the SSVB (We have provided further details pertaining to the spatio-spectral plane reconstruction in the supplementary information). The results shown in Fig.~\ref{fig:Experiment} (c) are that of the Spatio-spectral Stokes components and Stokes phases $\left(\Phi_{ij} = \arctan\left(\frac{S_j}{S_i}\right) \right)$ (shown as insets). 
It is clear from these results that $S_1$ and $S_2$ are dependent on the azimuthal coordinate oscillating between $1$ and $-1$ with a periodicity of $\frac{\pi}{2}$ as expected from Eq.~\ref{eq:SpatSpecStokesFP}, with a sign flip occurring whenever $f(\lambda)$ has cycled through $\approx\frac{\pi}{2}$. $S_3$ on the other hand solely depends on wavelength, exhibiting a sign flip after $f(\lambda)$ has cycled through $\frac{\pi}{2}$. Furthermore, we also note the emergence of  C-point 
polarization singularities embedded within $\Phi_{12}$  
indicating the presence of points of undefined polarization orientation and pure circular polarization. A full summary of the state-of-polarization (SOP) is shown in Fig.~\ref{fig:Experiment} (d) depicting the change in ellipse orientation with azimuthal angle and ellipticity change with wavelength. At particular wavelengths (the boundary between circular and elliptical states) we observe a $90^{\circ}$ flip in the orientation of the polarization ellipse, consistent with the sign flip observed in the Stokes parameters shown in Fig.~\ref{fig:Experiment} (c). This is due to tracing out lines of latitude on the Poincar\'e sphere with changing wavelength. \\

\noindent We note that in the sections to follow, every representation of the Stokes vector, such as Stokes parameter plots, SOP plots and vector texture plots, will be derived from the spatio-spectral Stokes vector, $\vec{S}(\lambda,\phi)$. \\

\begin{figure*}[t!]
\includegraphics[width=\linewidth]{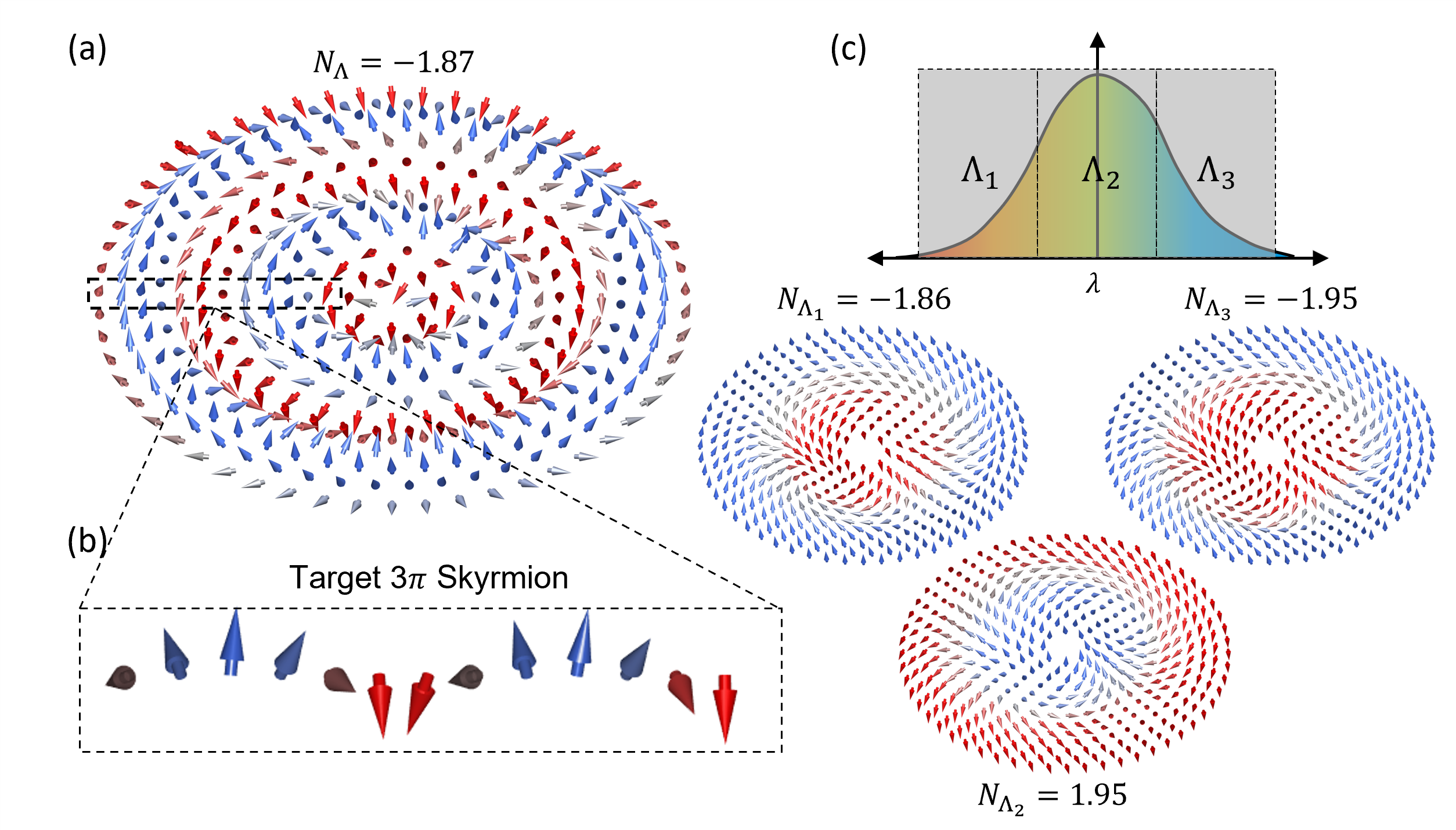}
\caption{\textbf{Spatio Spectral topology} (a) The full vector texture of the spatio-spectral Stokes revealing the entire topological structure of interest, characterized by a topological number $N_\Lambda = -1.87$. The vector texture is given by a vector pointing along the surface of the Poincar\'e sphere. (b) Tracing the vector field through wavelength reveals that the vector field rotates from the north to south pole 3 times. Therefore our topological structure is that of a $3\pi$ Skyrmion, an example of a target Skyrmion. (c) By partitioning the wavelength bandwidth of the field into three partitions $\Lambda_1,\Lambda_2,\Lambda_3$, we observe that each partition has a well-defined topology with topological observables that sum to give the topological invariant of the entire field.}
\label{fig:SpatSpecTopology}
\end{figure*}

\noindent \textbf{Topology of spatio-spectral vector beams.} We now consider the topology embedded within the spatial, spectral and polarization correlations. As previously established, the SSVB given in Eq.~\ref{eq:SpatSpecStokesFP} admits the mapping $\vec{S}(\lambda,\phi):R_{\Lambda}^2\xrightarrow{N_{\Lambda}} S^2$ where the spatio-spectral plane, $R_{\Lambda}^2$ wraps the Poincar\'e sphere, $N_{\Lambda}$ times. Adapting the typical Skyrmion number calculation by replacing the radial coordinate with the wavelength coordinate allows us to evaluate the topological invariant classifying our state as 

\begin{equation}
N_{\Lambda} = \frac{1}{4\pi} \int_{\Lambda} \int_0^{2\pi} \vec{S}(\lambda,\phi)\cdot \left(\partial_{\lambda}\vec{S}(\lambda,\phi) \times \partial_{\phi}\vec{S}(\lambda,\phi) \right) d\lambda\, d\phi ,
\label{eq:SpatSpecTop}
\end{equation}

where $\Lambda=[\lambda^-,\lambda^+]$ is the spectral domain over which we evaluate the Skyrmion number. Evaluating the Skyrmion number for the Stokes vector given by Eq.~\ref{eq:SpatSpecStokesFP} yields

\begin{equation}
N_{\Lambda} = \frac{1}{4\pi} [\cos(2f(\lambda))]_{\lambda^-}^{\lambda^+} \; [2g(\phi)]_0^{2\pi} = p_{\lambda}v_{\phi},
\label{eq:SpatSpecTopFBExp}
\end{equation}

where the term $p_{\lambda}=\frac{1}{2}[\cos(2f(\lambda))]_{\lambda^-}^{\lambda^+}$ is defined as the polarity and  $v_{\phi} =\frac{1}{2\pi}[2g(\phi)]_0^{2\pi} = 4q $ the vorticity of the vector field. Computing the Skyrmion number over the entire spectral bandwidth yields $N_{\Lambda} = -1.87$, consistent with a theoretically predicted Skyrmion number of $2$. We can study the topological structure more closely by plotting the mapping as a vector texture, as shown in Fig.~\ref{fig:SpatSpecTopology} (a). Here the vector is seen to rotate about the $S_3$-axis with a change in azimuthal coordinate and oscillate between the north pole and south pole with a change in the spectral coordinate. Taking a cross-section of the vector field, as shown in Fig.~\ref{fig:SpatSpecTopology} (b), reveals that the vector undergoes three full $\pi$ rotations from the north to south Pole of the Poincar\'e sphere within the full spectral domain signifying that our SSVB is characterized by a Target $3\pi$ Skyrmion \cite{shen2023optical}. A signature of a $k\pi$ target Skyrmion is that it can be decomposed into $k$ individual Skyrmions. 
Following this convention, our SSVB admits a decomposition into three distinct Skyrmionic vector fields, achieved through partitions of the spectral domain according to $\Lambda_{1} = [1546,1554]\text{nm},  \; \Lambda_{2} = [1554,1563]\text{nm}, \; \Lambda_{3} = [1563,1571]\text{nm}$, as shown in Fig.~\ref{fig:SpatSpecTopology}(c). Each vector field can be characterized by an integer-valued Skyrmion number calculated to be $N_{\Lambda_{1}} = -1.86, \; N_{\Lambda_{2}} = 1.95, \; N_{\Lambda_{3}} = -1.95$, which evidently sum to give the result obtained for the full vector field. \\

\noindent Having identified the topology embedded within the mutual spatial, spectral and polarization correlations, we now explore the tunability of this spatio-spectral Skyrmionic texture.\\

\begin{figure}[t!]
\includegraphics[width=\linewidth]{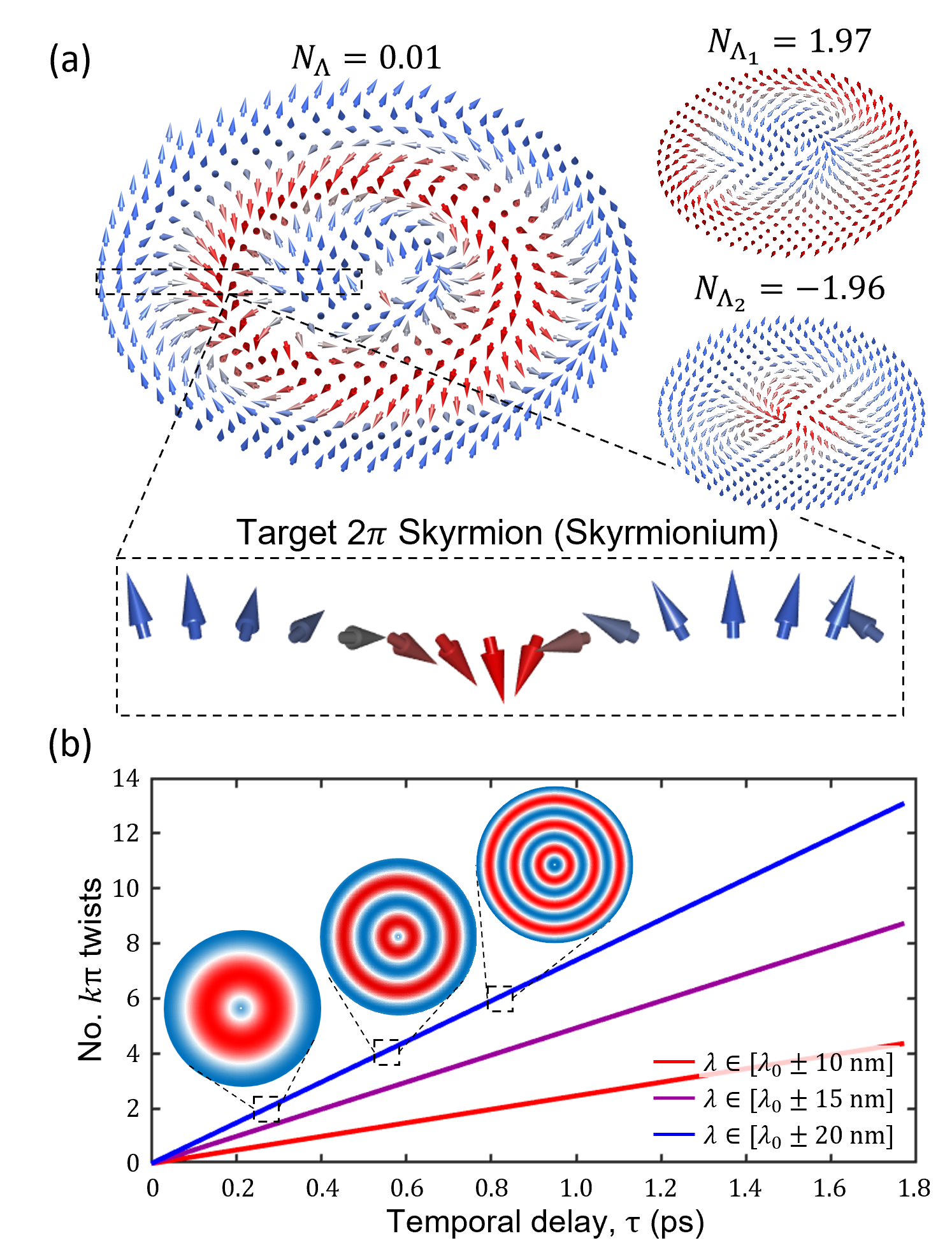}
\caption{\textbf{Tunable Target topology.} (a) Experimentally measured vector texture of Spatio-spectral Skyrmionium (left) with composite topological structures (right). (b) Numerical simulation of number of $\pi$ rotations of the field given as a function of temporal delay imparted by the BBO within different spectral domains with $S_3$ shown as an inset.}
\label{fig:TopKpiRot}
\end{figure}

\noindent \textbf{Tunable $k\pi$-twist structure.}
In the previous result, we have shown that the Skyrmion possessed a $3\pi$-spectral twist structure, however this parameter can be manipulated by tuning the delay of the formative pulses of the SSVB. For a $k\pi$-twist within the spectral domain, $\Lambda = [\lambda^-,\lambda^-]$, the temporal delay required is given by $\Delta t = k\frac{\lambda^+\lambda^-}{4c (\lambda^+ - \lambda^-)}$. By replacing one of the $2$ mm crystals with a $0.5$ mm crystal (See supplementary for more details) we achieved a Skyrmion structure with a $2\pi$-twist, also known as a Skyrmionium with the results shown in Fig.~\ref{fig:TopKpiRot} (a). Here the total Skyrmion number evaluates to $N_{\Lambda} = 0.01$ close to the expected value of zero, with the cross-section of the field shown as an inset. Partitioning the field into its constituent components reveals the existence of two Skyrmions with opposite polarities, thus resulting in a sign difference between their Skyrmion numbers, $\{N_{\Lambda_1},N_{\Lambda_2}\} = \{1.97,-1.96\}$ which sum to give the Skyrmion number of the total field. 
Fig.~\ref{fig:TopKpiRot} (b) shows the linear relationship between the temporal delay and number of spectral-twists with $S_3$ shown as insets for particular temporal delays. Additionally, we note that by increasing the spectral domain within which we measure our SSVB, we also naturally observe more spectral twists as shown in Fig.~\ref{fig:TopKpiRot} (b) for the spectral domains $[\lambda_0\pm10 \text{nm}$, $[\lambda_0\pm15 \text{nm}]$ and $[\lambda_0\pm20 \text{nm}]$. Albeit we note that to reach larger spectral domains in practice would require shorter pulses than that used in the current experiment. Nevertheless, the control over the temporal delay results in more spectral rings seen in $S_3$ within a given spectral domain, thereby enabling control over the number of spectral twists exhibited by the spatio-spectral Skyrmion structure.\\

\noindent \textbf{Topological multiplexing.} 
Up until now we have omitted the radial DoF of the transverse spatial domain of the light field. We now consider exploiting this additional dimension in order to construct an SSVB with radially dependent topology, i.e., we implement a topological multiplexing. We achieve this by passing the SVB through an ``O-plate"; a modified s-plate with a radially dependent charge, $q(r)$, as shown in Figure~\ref{fig:TopMulti} (a), (further details regarding the O-plate design and manufacture are given in the supplementary information). The generated SSVB is expressed as an extension of Eq.~\ref{eq:FieldTransform1} as follows 
\begin{equation}
|\Psi\rangle = \cos(f(\lambda)) e^{ig(r,\phi)}|R\rangle + \sin(f(\lambda)) e^{-ig(r,\phi)}|L\rangle,
\label{Eq:MultiTop}
\end{equation}
where $g(r,\phi) = 2q(r)\phi$ is a radially-dependent function encoding different integer charges at different non-overlapping radial regions, $R_i$. Experimental results are shown in Fig.~\ref{fig:TopMulti} for the case where 3 different Skyrmion numbers, $q(r)\in\{\frac{1}{2}, 1, \frac{3}{2}\}$ are encoded within three distinct radial regions, $R_1,R_2$ and $R_3$, such that 
\begin{equation}
    g(r,\phi) =\begin{cases} 
-\phi & \text{if \;} r \in R_1, \\
-2\phi & \text{if \;} r \in R_2, \\
-3\phi & \text{if \;} r \in R_3. \\
    \end{cases}\\
\end{equation}

The state given in Eq.~\ref{Eq:MultiTop} can be written as a linear superposition of three orthogonal states, $|\psi\rangle = | \psi_1 \rangle + | \psi_2 \rangle + | \psi_3 \rangle$ where

\begin{equation}
    | \psi_i \rangle = \cos(f(\lambda)) e^{ig(R_i,\phi)\phi}|R\rangle + \sin(f(\lambda)) e^{-ig(R_i,\phi)\phi}|L\rangle.
\end{equation}

with the fact they are defined over non-overlapping regions ensuring their orthogonality, $\langle\psi_i|\psi_j\rangle = \delta_{ij}$. Since each state does not overlap spatially with any other state, their topological numbers can be controlled independently. In this way the wavelength and azimuthal degrees of freedom are used to define a full mapping to the Poincar\'e sphere, with the topological wrapping number controlled by the radial degree of freedom, encoded through the function $g(r,\phi)\in \mathbb{Z}$. Therefore, this scheme allows for the encoding of multiple Skyrmion numbers within a single SSVB thereby achieving an increase in information capacity through topological multiplexing. In Fig.~\ref{fig:TopMulti} (b) the different Skyrmion encoding levels of our SSVB are shown. We perform several non-overlapping radial projections of our SSVB within each radial ROI, $R_i$, and observe that the Skyrmion number within each regime remains constant, only changing upon a transition between ROIs, with experimental measurements matching numerical simulations. We note that the numbers were calculated within a single twist of the skyrmion structure. Example topological structures for 3 distinct radial projections are shown in Fig.~\ref{fig:TopMulti} (c) with calculated Skyrmion numbers $N=\{1.91, 3.90,5.8\}$. The insets show the change in the vector textures for each radial projection thereby emphasizing the change in vorticity between each radial ROI, $v_{\phi} = 2,4,6$ in $R_1,R_2,R_3$, respectively.
While this proof of principle demonstration shows the encoding of 3 topological numbers in a single SSVB, this scheme is scalable allowing for the encoding of an arbitrary number of topological numbers within a single SSVB, limited only by the aperture of the system. 

\begin{figure*}[t!]
\includegraphics[width=\linewidth]{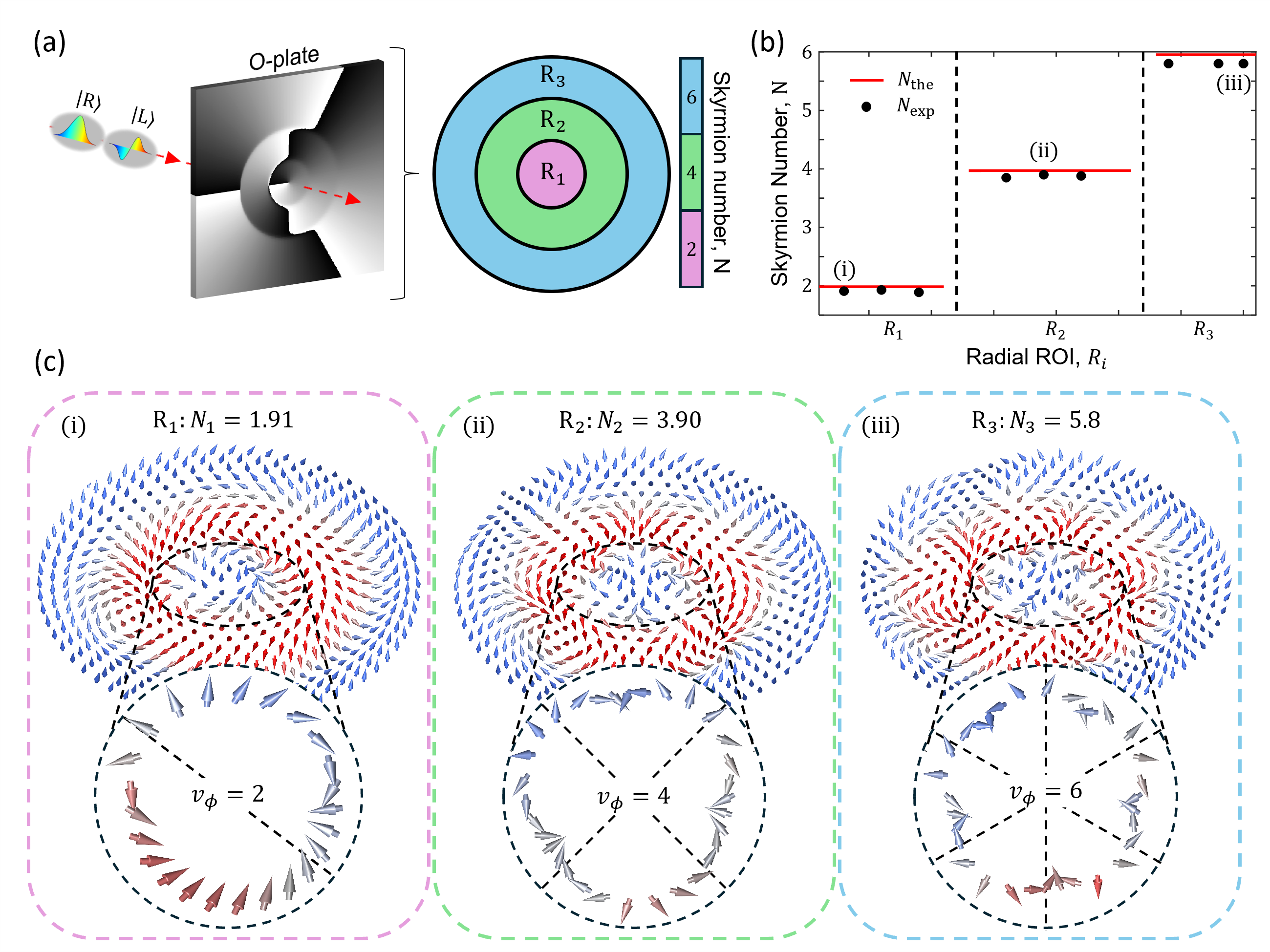}
\caption{\textbf{Topological multiplexing.} (a) An SVB incident on an O-plate with a radially dependent charge thereby encoding different topological numbers at different radial regions, $R_i$. (b) Line plot depicting the calculated Skyrmion numbers for a simulated (red line) and experimental (black points) SSVB within different radial regimes. 
(c) Vector textures extracted from radial projections (i - iii) within distinct radial encoding regimes, $R_i$, with insets emphasizing the change in vorticity observed in the fields azimuthal coordinate.} 

\label{fig:TopMulti}
\end{figure*}

\section*{DISCUSSION and CONCLUSION}

\noindent We have reported the first spatio-spectral target Skyrmion derived from the non-separability between the spectral, spatial and polarization DoFs of a light pulse. Using a simple and compact generation technique, consisting of only a birefringent crystal and vortex plate, we have encoded the desired Skyrmionic structure into our light field. By representing the tripartite-like correlations embedded within the state as a map between the spatio-spectral plane and the Poincar\'e sphere we have revealed the target Skyrmion configuration consisting of a spectral-$k\pi$-twist structure that could be decomposed into $k$ Skyrmions with Skyrmion numbers that add to give that of the total topological structure. Furthermore, we demonstrated control over the number of spectral-twists through tuning of the temporal delay of the formative pulses of the SSVB revealing an additional tunable parameter. Lastly, we have proposed a novel multiplexing scheme which exploits all dimensions (2 space + 1 spectral) of our SSVB by independently encoding different Skyrmion numbers at different radii of the field. We demonstrate the practicality of this approach by realizing an SSVB characterized by three distinct Skyrmion numbers ($N=1.91, 3.90, 5.80$). This work opens up a new route towards dense information encoding using multiple topological channels encoded in a single light field. 

\noindent In this work we have focused on a simple, compact proof-of-principle experiment which demonstrated the ability to generate complex SSVBs exhibiting highly tunable Skyrmion topological features. However, the scheme we have outlined can be easily generalized using dynamic modulation techniques. For example, by replacing the BBO and s-plate with a Mach-Zehnder interferometer, it would be possible to dynamically change the path length of one of the polarization arms in order to dynamically alter the number of $k\pi$ twists exhibited by the topology. Additionally, an SLM placed within the interferometer would allow for the dynamic modulation of the amplitude profile of the field thus allowing for a more general topological encoding scheme with a tunable number of encoded topological numbers within a single field. Furthermore, in contrast to typical optical Skyrmions embedded in the spatial domain \cite{gao2020paraxial}, our scheme achieves Skyrmions of varying topological number with spatial modes of the same mode order. Thus, ensuring that the field is propagation-invariant due to the identical accumulation of mode-dependent Gouy phases in both modes.
Similarly, it will be interesting to explore the resilience of spatio-spectral Skyrmions to noise and environmental disturbances as the topology is spanned across two different domains possibly increasing the known protection mechanism.
Lastly, sending the SSVB through a dispersive medium, such as a long fiber, would map the spectral features onto the temporal DoF, thereby providing a novel pathway for the creation of tunable spatio-temporal Skyrmions without the need for complex experimental setups \cite{zhan2024spatiotemporal}. \\

\noindent In conclusion, we believe that this work extends the optical toolkit for optical topological design beyond the spatial \cite{shen2024optical} and spatio-temporal \cite{teng2025construction} domains. Furthermore, this work may inspire new directions into embedding diverse topological structures, such as polarization singularities \cite{dennis2009singular,  cardano2013generation, otte2018polarization}, links, knots \cite{larocque2018reconstructing}, hopfions \cite{shen2023topological, wan2022scalar} and Skyrmionic-hopfions \cite{sugic2021particle, ehrmanntraut2023optical}, etc. into the spatio-spectral domain of optical fields. We believe that our approach to topological multiplexing may further inspire future investigations into optimal topological encoding schemes which seek to maximize the topological information packed into a single optical field. 

\section*{Acknowledgements}
The authors would like to thank Lea Kopf, Jaime Moreno, Piotr Ryczkowski, Uttam Kumar Samanta, Jiaqi Li for their help and useful discussions about the experiment. This work was supported by the South African National Research Foundation, National Institute for Theoretical and Computational Science and CSIR Rental Pool Programme. P.O acknowledges funding from the Council for Scientific and Industrial Research under the HCD-IBS scholarship scheme. NM acknowledges the financial support from the Photonics Research and Innovation Flagship (PREIN - Decision 346511). OK and RF acknowledge the support of the Research Council of Finland through the project BIQOS (decision 358134). RF acknowledges the support of the Research Council of Finland through the Academy Research Fellowship (decision 332399) and  European Research Council (ERC) Starting grant TWISTION (101042368).

\section*{Author contributions}
The idea was conceived by PO, IN and RF. The experimental setup was designed by PO and NM. The ``O-plate" design, fabrication, and calibration was performed by OK. The measurements and data analysis was performed by PO and NM. The code for the spatio-spectral Stokes reconstruction was written by PO. PO and NM wrote the manuscript with revisions and edits from OK, IN, AF and RF. RF, IN and AF supervised the project.

\section*{Competing Interests}
The authors declare no competing interests.

\section*{Data availability}
The data are is available from the corresponding author on request.

\clearpage
\appendix

\setcounter{section}{0}
\setcounter{figure}{0}
\setcounter{table}{0}
\setcounter{equation}{0}
\setcounter{footnote}{0}
\renewcommand{\thesection}{S\arabic{section}}
\renewcommand{\thefigure}{S\arabic{figure}}
\renewcommand{\thetable}{S\arabic{table}}
\renewcommand{\theequation}{S\arabic{equation}}

\begin{widetext}

\section*{Supplementary: Experimental generation and measurement of spatio-spectral skyrmions}

\noindent Here we discuss specific details pertaining to the generation, characterization and measurement of spatio-spectral vector beams (SSVB) with tunable topological features. A schematic of the detailed experimental setup is shown in Fig.\ref{fig:supp_FullExp}.

\subsection{State Generation} 
\noindent A femtosecond laser with a 1560 nm ($\pm$5 nm) central wavelength, 220 fs pulse duration, and 80 MHz repetition rate is used as a source (see Fig. \ref{fig:supp_FullExp}).The pulsed laser's polarization is adjusted using a polarizing beam splitter (PBS) and a set of waveplates (quarter-wave plate Q1, half-wave plate H1). Two consecutive birefringent BBO (BBO1/2) crystals are used to coherently split the pulse into two temporally separated, orthogonal polarization components corresponding to the crystals' ordinary and extraordinary axes. 
Different combinations of BBO crystals are used to tune the group delay between the split pulses, with further details provided in the proceeding section. 
Finally, a vortex half-wave retarder (S-plate) couples the polarization to the spatial degree of freedom, creating an SSVB. 
Furthermore, by adjusting the group delay imparted by the pair of BBOs, we are able to tune the number of spectral-twists exhibited by the SSVB within its spectral domain, as shown in the main text (Fig. 4). 

\subsubsection{Tunable group delay imparted by BBOs}

\noindent Different BBO configurations are used in the experiment with the goal of tuning the total imparted relative group delay $\Delta\tau$ between the two orthogonally polarized pulses.\\

\noindent The delay between the two pulses in a BBO crystal is calculated from the wavelength dependent group indices of ordinary ($o$) and extraordinary ($e$) components. First, the wavelength ($\lambda$) dependent refractive indices $n_o(\lambda)$ and $n_e(\lambda)$ are obtained from the Sellmeier equations of the BBOs \cite{tamovsauskas2018transmittance}. The group index is connected to the refractive index as $n_g(\lambda) = n(\lambda) - \lambda \, dn/d\lambda$. For a cut angle $\theta$, the effective group index of the extraordinary component is calculated as
\begin{equation}
n_{g,\text{eff}}(\lambda,\theta) = \frac{n_{g,o}(\lambda)\, n_{g,e}(\lambda)}{\sqrt{n_{g,o}^2(\lambda)\sin^2\theta + n_{g,e}^2(\lambda)\cos^2\theta}}.
\end{equation}
Finally, the relative group delay between the ordinary and extraordinary components for a crystal of thickness $d$ is written as
\begin{equation}
\Delta \tau(\lambda,\theta) = \frac{d}{c}\,\left(n_{g,o}(\lambda) - n_{g,\text{eff}}(\lambda,\theta)\right),
\end{equation}
where $c$ is the speed of light in vacuum.\\ 

\noindent In the experiment, two different dual BBO configurations were used with different thicknesses and cut angles: $\{d_1,d_2,\theta_1,\theta_2\}=\{2 \text{mm}, 2 \text{mm},41.8^\circ, 23.4^\circ\}, \{2 \text{mm}, 0.5 \text{mm},41.8^\circ, 29.8^\circ\}$. We note that the dual BBO configuration allowed for mitigation of spatial walk-off effects by orienting the two crystals $180^{\circ}$ with respect to one another. The first configuration imparts a delay of $\Delta\tau = 0.48$ ps while the second imparts a delay of $\Delta\tau = 0.4$ ps. An increasing delay leads to faster oscillation (evolution) of the polarization in the spectral domain. Therefore, the control over the temporal delay between the constituent split pulses provides an avenue for the tuning of the number of full $k\pi$ rotations exhibited by the Skyrmionic mapping within the spectral domain of the SSVB. 

\subsection{Measurement} 
\noindent The SSVBs are nonseparable states involving three DoFs: spatial, spectral, and polarization. Therefore, to reveal their topological features in experiments, measurements are performed simultaneously across all three DoFs (see Fig. \ref{fig:supp_FullExp}). First, a specific polarization state is selected using waveplates  (H2, Q2) and a polarizing beam splitter (PBS2). The spectral components are then filtered with a home-built monochromator, consisting of a blazed grating of 600 lines/mm grooving, two planoconvex lenses L3, L4 and a spectral slit placed at the focal plane of L3. Finally, the spatial profile of the polarization- and frequency-filtered light field is recorded with a camera.

\begin{figure*}[t!]
\includegraphics[width=\linewidth]{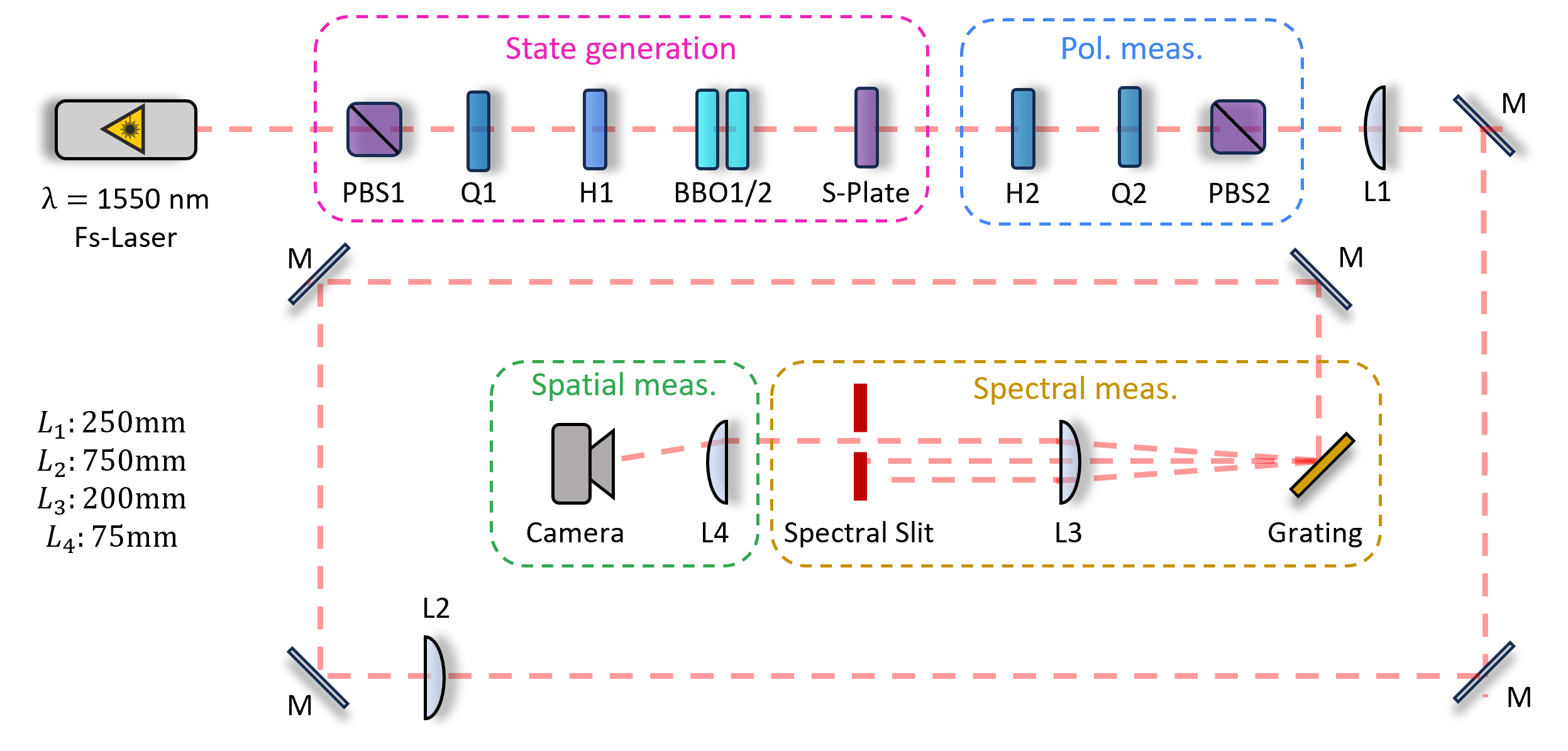}
\caption{\textbf{Experimental setup to generate and characterize SSVBs}. Q: quarter waveplates, H: half waveplates, BBO: Beta Barium Borate crsytals, S-Plate: vortex half wave retarder, PBS: polarizing beam splitter, M: mirrors, L: lenses. The SSVB is measured by measuring all three degrees of freedom: polarization, spectrum and space respectively. 
}
\label{fig:supp_FullExp}
\end{figure*}

\subsubsection{Spectral vector beam characterization}
\noindent To correctly generate and measure the SSVB, we first need to characterize the spectral-polarization correlations embedded in the SVB that exists before passing the field through the S-plate. To characterize the SVB in our system, polarization-resolved intensities are captured by placing the camera at the Fourier plane (back focal plane of L3) of the grating to directly record the spectral domain of the pulse. The different SVB results are shown in Fig.~\ref{fig:supp_SVBCharacterization}. This characterization additionally provides a calibration of the spectral plane for future spectral measurement of the SSVB (see proceeding section).\\ 
 
\begin{figure*}[t!]
\includegraphics[width=\linewidth]{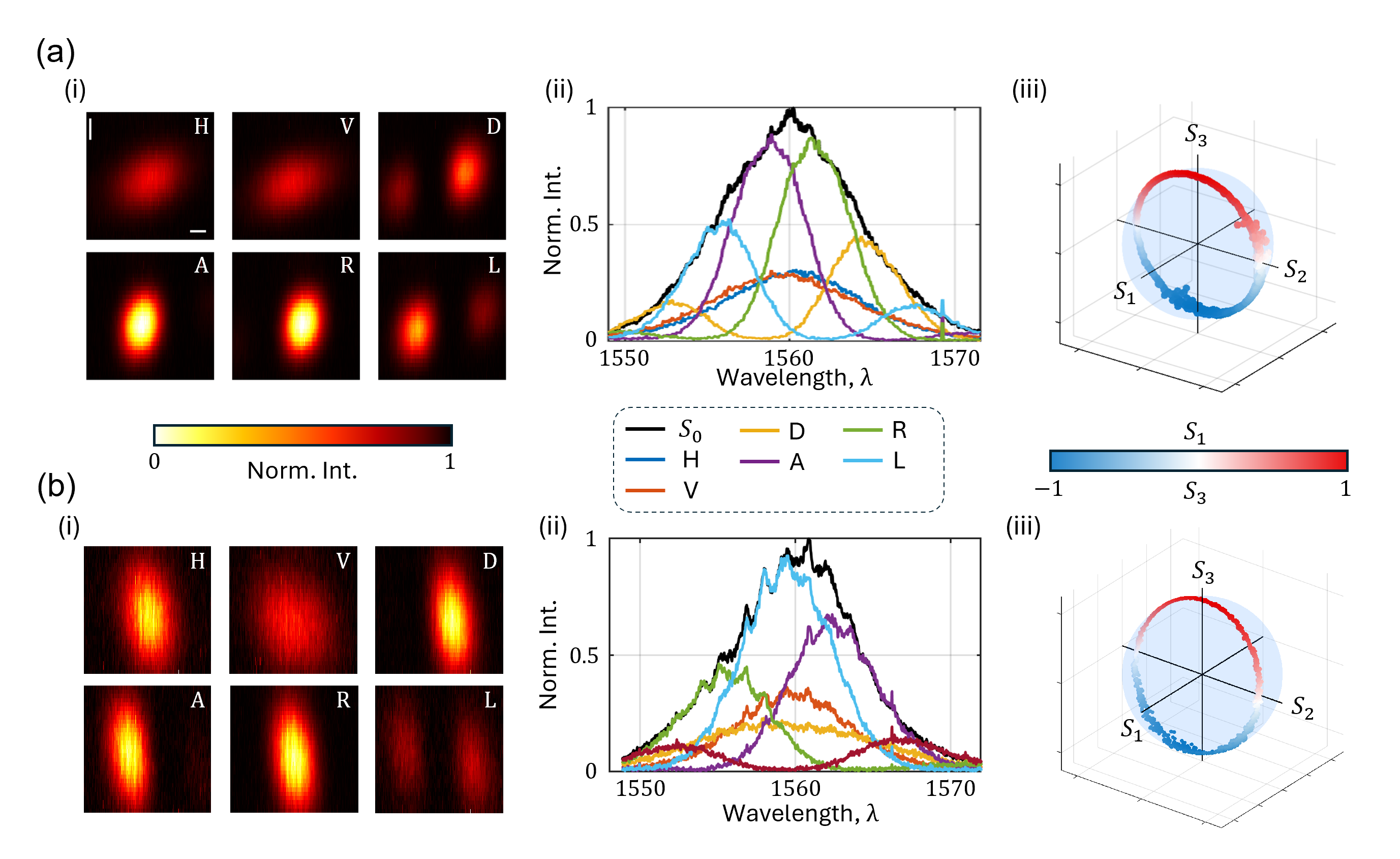}
\caption{\textbf{Experimental characterization of spectral vector beams (SVB)} 
Results for SVB generated in the $|H\rangle,|V\rangle$ basis, using a (a) 4mm and (b) 2.5mm dual BBO configuration, respectively. 
}
\label{fig:supp_SVBCharacterization}
\end{figure*}

\noindent In the experiment we generate two different SVBs, with the first generated using a 4 mm thick dual BBO configuration ($\{d_1,d_2,\theta_1,\theta_2\}=\{2 \text{mm}, 2 \text{mm},41.8^\circ, 23.4^\circ\}$) 
and the second using a 2.5 mm thick dual BBO configuration ($\{d_1,d_2,\theta_1,\theta_2\}=\{2 \text{mm}, 0.5 \text{mm},41.8^\circ, 29.8^\circ\}$). The spectral polarization evolution for each configuration is measured through six standard polarization projections $\{H, V, D, A, R, L\}$ with simultaneous spectral measurements performed by the grating, L3 and camera. The camera images at the spectral plane for all six polarization projections are presented in Fig. Fig.~\ref{fig:supp_SVBCharacterization}(a-b) (i). The total intensity $S_0$ is then fitted with a spectral profile of the laser source to get a map from camera pixel to wavelength. The spectral line profiles for different projections are shown in Fig.~\ref{fig:supp_SVBCharacterization}(a-b) (ii). Since the grating is orientated to spread the wavelength horizontally, the spectral line profiles are obtained by averaging over the rows of pixels on the camera images. Lastly, to highlight the nature of the spectral-polarization correlations, the reconstructed spectral-dependent Stokes vectors, $\vec{S}(\lambda) = (S_1 \; S_2 \; S_3)^T$, are plotted on the Poincar\'e sphere as shown in Fig.~\ref{fig:supp_SVBCharacterization}(a-b) (iii). 
In this configuration, the input pulse polarization is set to diagonal by H1, and the BBO splits it into two pulses with horizontal and vertical polarization. 
Here, the polarization evolves across the beam's spectrum according to $|P\rangle \propto |H\rangle+e^{if(\lambda)|V\rangle}$, with no relative amplitude change between $|H\rangle$ and $|V\rangle$ as seen by their equal intensity projections in Fig.~\ref{fig:supp_SVBCharacterization}(a) (i). More specifically by inspection of the spectral projections shown in Fig.~\ref{fig:supp_SVBCharacterization}(a) (ii) the polarization of the SVB evolves as $|R\rangle \to |D\rangle \to |L\rangle \to |A\rangle \to |R\rangle \to |D\rangle \to |L\rangle$ across the spectral domain of the SVB. Therefore the polarization evolves about the $S_1$ axis, performing approximately a $3\pi$ rotation around the $S_1$ axis across the spectral domain. 
In the second case, the shorter 2.5 mm thick dual BBO configuration was used with the resulting measurements shown in Fig. \ref{fig:supp_SVBCharacterization}(c). In this configuration the generated SVB exhibits similar correlations as that exhibited by the SVB in the previous case except that the shorter delay induced by the shorter dual BBO configuration results in a $2\pi$ rotation about the $S_1$ axis.\\

\subsubsection{Full characterization of spatio-spectral vector beams}

\noindent In the preceding section, we analyzed 2 different SVBs used as inputs into s-plate used to generate the final desired SSVB. 
In this work the two SVB cases considered were used to generate  SVB with a $3\pi$ and a $2\pi$ spectral twist, respectively, with the latter used to form a Skyrmionium structure. Here we discuss practical considerations for the measurement of the SSVBs with the results shown for the first SVB passed through the s-plate. \\ 

\noindent As discussed in preceding sections, the measurement of the SSVB involves performing sequential projections onto polarization, wavelength and space. This was done using the measurement schematic depicted in Fig.~\ref{fig:supp_SSVBData} (a). We note here that the spectral and spatial measurements must be performed with care. In the previous section, spectral measurements were performed directly on the camera, therefore the combination of the grating constant, the focal length of L3 and pixel size sets the spectral resolution, $\Delta \lambda_{res}$. By fitting a spectral function to $S_0$ at the focal plane of L3, we find that the full width half maximum (FWHM) of the spectrum spans $1.22$ mm on the camera sensor. Dividing by a camera pixel size of $3.45 \mu m$ and multiplying by a FWHM of $\Delta \lambda = 10$ nm gives an effective spectral resolution of $\Delta \lambda_{res} = 28.3$ pm for our SVB. However, in the case of measuring the full SSVB we note that a spatial measurement is performed twice, first as part of our spectral projection (via our spectral slit within our home-made monochromator) and second as part of our spatial measurement performed by the camera. Therefore, the size of the SSVB in the far-field field limits the spectral resolution of our measurement. In our experiment, the spectral slit was set to a slit size of 0.5mm translating to a spectral resolution of $\Delta \lambda_{res} = 4.10$ nm. Since this is relatively large compared to the FWHM we shifted the slit by 0.1 mm (corresponding to a spectral shift of around 0.82 nm) for each measurement in order to acquire a larger set of measurements. To ensure that we obtained the expected behaviour despite the diminished spectral resolution, we calibrated the spatio-spectral Stokes parameter, $S_3(\vec{r},\lambda)$, obtained for the SSVB measurement against that obtained for the SVB measurement, $S(\lambda)$. The results are shown in Fig.~\ref{fig:supp_SSVBData} (b) with the SSVB data given as purple points and the SVB data shown as a blue line. The behaviour of the experimental data for the SSVB measurement matches the behaviour of that measured in the SVB measurement therefore demonstrating that the diminished spectral resolution does not impact the observed spectral-spatial-polarization correlations. This is expected since the reduced spectral resolution results in a diminished degree of polarization. However, this does not affect the orientation of the Stokes vectors, therefore after normalization the original Stokes vector is recoverable. We note that in the scenario that a higher spectral resolution is required, the grating constant can be increased in order to spread the spectrum over a larger spatial domain and furthermore the beam size before L3 can be increased in order to yield a smaller beam size at the focal point of L3 therefore allowing a smaller spectral slit size without ``clipping" the spatial mode in the far-field.
 Additionally, the expected behaviour of each polarization projection, recalling that our produced SSVB takes the form, $|\Psi\rangle = e^{ig(\phi)}\cos(f(\lambda))|R\rangle + e^{-ig(\phi)}\sin(f(\lambda))|L\rangle$, can still be easily discerned from the spatio-spectral-polarization projections shown in Fig.~\ref{fig:supp_SSVBData} (c). 
Lastly, the polarization population of our SSVB is depicted in Fig.~\ref{fig:supp_SSVBData} (d) indicating full coverage over all polarization states indicating that our SSVB exhibits the desired correlation between spectrum, space and polarization.

\begin{figure*}[t!]
\includegraphics[width=\linewidth]{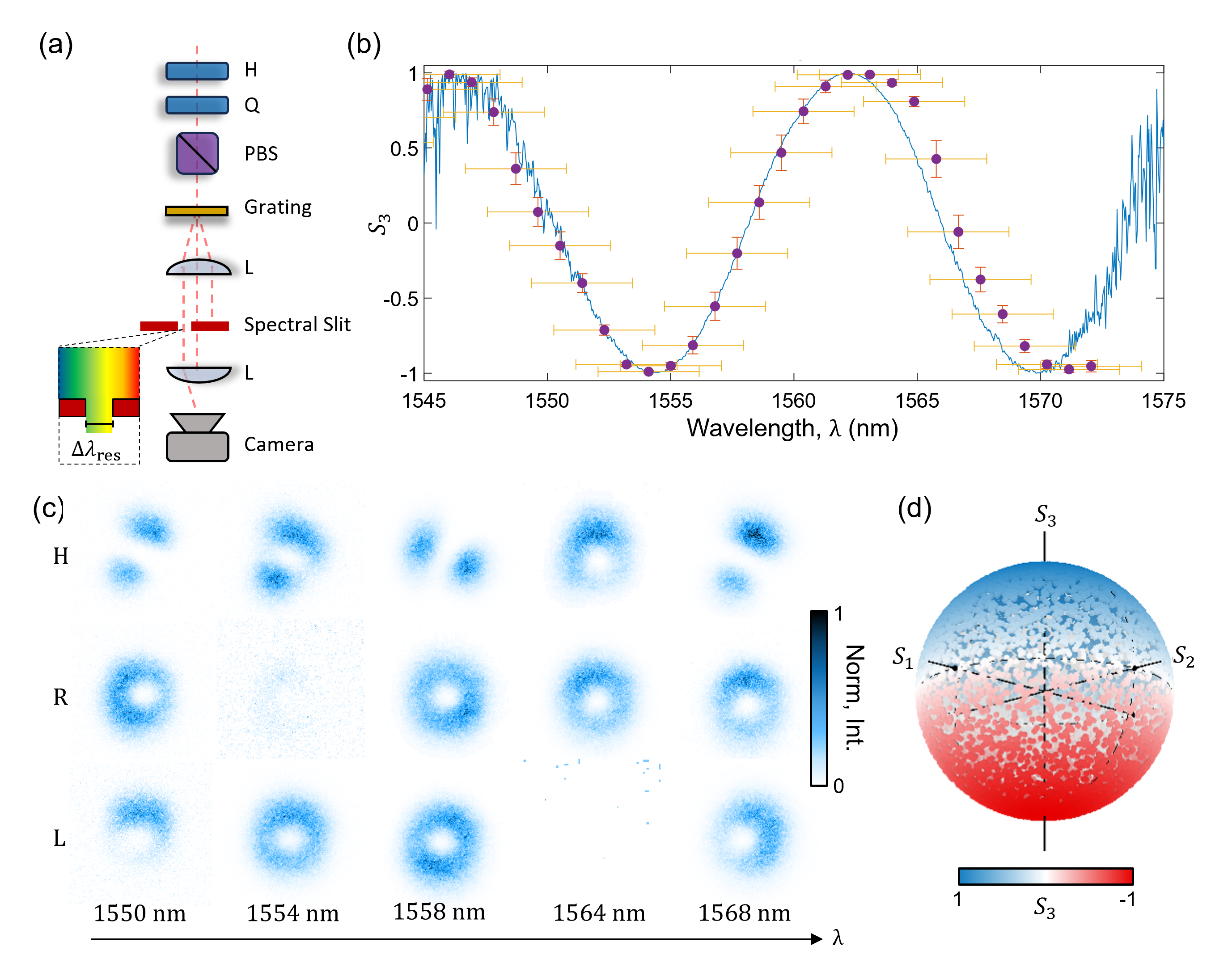}
\caption{\textbf{Experimental characterization of spatio spectral vector beams (SSVB)}. (a) Schematic of experimental setup used to simultaneously measure the polarization, spectral and spatial DoFs of the SSVB. (b) Experimentally measured normalized Stokes parameter, $S_3$, as a function of wavelength as measured for the SVB (blue line) and SSVB (purple dots). (c) Example spectral-polarization projections for the SSVB. (d) Poincar\'e sphere plot indicating polarization state population embedded within the spatio-spectral correlations.}
\label{fig:supp_SSVBData}
\end{figure*}

\section*{Supplementary: S-plate with radially structured topological charge, ``O-plate"}
\noindent The s-plate used in this work imparts the relative phase structure and ultimately controls the magnitude of the Skyrmion number, $N$. However, with the relative phase structure imparted azimuthally and the relative amplitude change encoded in the spectral bandwidth, the radial degree of freedom is left open to exploit for further purpose. In this work we opt to exploit the radial DoF for the multiplexing of topology, specifically different azimuthally-varying relative phases are encoded at differ radii. This transformation is implemented via a custom built spatially tailored birefringent meta-structure, we have named the ``O-plate". In the sections to follow we discuss its design, fabrication process and experimental characterization.

\begin{figure*}[t!]
\includegraphics[width=0.85\linewidth]{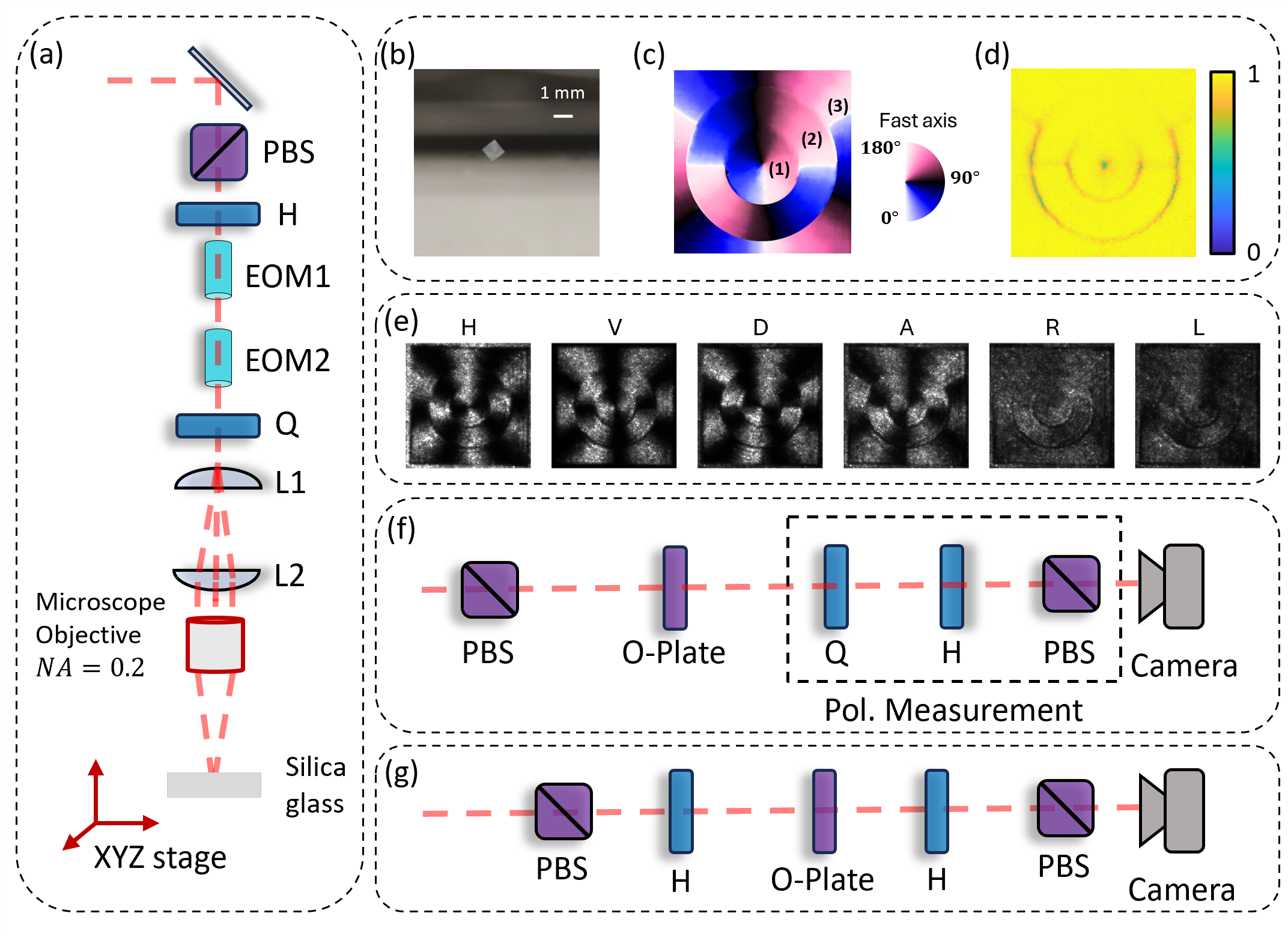}
\caption{\textbf{Fabrication process and characterization of O-plate.} (a), Schematic of the femtosecond laser writing setup used to fabricate the O-plate inside the glass substrate. (b), Photograph of the fabricated O-plate, showing the written structure and its physical size (scale bar: 1 mm). (c), Spatial distribution of the slow-axis orientation across the entire plate, highlighting the different radial regions (1–3). (d), Normalized transmission map of the O-plate, where a value of 1 corresponds to a half-wave retardance. (e), Intensity images corresponding to all polarization projection measurements (H, V, D, A, R, and L), used for polarization tomography. (f), Experimental setup for polarization tomography used to reconstruct all polarization projections. (g), Setup used for normalized transmission measurements. }
\label{fig:Oplate}
\end{figure*}

\noindent \subsection{O-plate Design}
The design of the O-plate is illustrated in Fig. \ref{fig:Oplate}. The orientation of the fast axis varies continuously from $0$ to $\pi$, resulting in a total polarization rotation of $2\pi$. This rotation differs across regions, with a total given by $2n\pi$, where $n=1,2,3$, as illustrated in  Fig. \ref{fig:Oplate}.(c). The local retardance is designed to be $\pi$ at the operating wavelength of $1550 \, \text{nm}$. However, small fluctuations in the retardance across the plate are observed experimentally, as shown in  Fig. \ref{fig:Oplate}(d).
\subsection{Fabrication method for O-plate}
The O-plate was fabricated using a technique known as direct ultrafast laser writing of nanogratings, which enables the precise inscription of sub-wavelength structures inside transparent materials by focusing ultrashort laser pulses \cite{sakakura2020ultralow,della2008micromachining,sugioka2014ultrafast}.

To achieve the required design, self-assembled nanogratings were inscribed in silica glass using a femtosecond fiber laser operating at 1030 nm, with a repetition rate of 251 kHz and a pulse duration of 209 $fs$. These values correspond to the operating parameters used during fabrication. The laser beam was focused into the sample by a microscope objective with a numerical aperture (NA) of 0.2, resulting in a focal spot size of approximately 2$\sim$3 $\mu m$. The silica glass substrate was mounted on a 2D translation stage to control the scanning position and speed. The orientation of the nanograting slow axis was controlled by the polarization of the writing laser, which was modulated by a polarizing beam splitter (PBS), a half-wave plate (H), two synchronized electro-optic modulators (EOMs), and a quarter-wave plate (Q). The laser writing setup is illustrated in Fig. \ref{fig:Oplate}.(a)

The O-plate was fabricated by writing parallel nanograting lines into the glass sample, separated by 1~$\mu$m, forming an area of 1~mm~$\times$~1~mm. The sample was translated at a speed of 0.2~mm/s, with a pulse energy of 1.2~$\mu$J, while the voltages were applied simultaneously to the EOMs according to the design of the O-plate structure. To achieve a local retardance of $\pi$, optimized for operation at 1550~nm, two nanograting layers with $\pi/2$ retardance were inscribed, separated by 100~$\mu$m.

\subsection{Characterization of the O-plate}

The fabricated O-plate was characterized using two measurements.
\paragraph{Slow axis orientation measurement (polarization tomography,  Fig. \ref{fig:Oplate}.(f)).}
A linearly polarized beam was transmitted through the O-plate, and polarization tomography was performed using a quarter-wave plate, a half-wave plate, a polarizing beam splitter, and a camera. Six polarization projections were recorded (H, V, D, A, R, L) using a camera, from which the Stokes parameters $(S_0, S_1, S_2, S_3)$ were calculated:
\[
S_0 = I_H + I_V, \quad 
S_1 = I_H - I_V, \quad 
S_2 = I_D - I_A, \quad 
S_3 = I_R - I_L,
\]
where $I_H, I_V, I_D, I_A, I_R, I_L$ are the measured intensities in the respective polarization bases as shown in Fig. \ref{fig:Oplate}.(e).  
The orientation angle of the slow axis $\theta$ was deduced as
\[
\theta = \frac{1}{2} \arctan\!\left(\frac{S_2}{S_1}\right).
\]
Since $\arctan$ returns values only in the range $[-\pi/2, \pi/2]$, phase unwrapping must be applied numerically to reconstruct a result in a range of $[0, \pi]$

\paragraph{Retardance measurement (Fig. \ref{fig:Oplate}.(g)).}
To measure the local retardance, the O-plate was placed between two polarizers, each consisting of an half-wave plate and a PBS. Normalized transmission $T_n(\theta)$ was measured as a function of angle $\theta$, giving the expected sinusoidal modulation:

\[
T_n(\alpha) = \frac{1}{2}\sin^2\!\left(2\theta\right) [1-\cos(\delta)],
\]

where $\delta$ is the local retardance of the O-plate, and $\theta$ is the angle between its fast axis and the first polarizer. A local maximum of the normalized transmission of 1, i.e., $T_n^{max}= 1$, indicates that the O-plate retardance at that location is $\pi$.

\section*{Supplementary: Spatio-spectral Stokes construction}

\noindent As discussed in the previous sections, the polarization profile of the SSVBs is evaluated at every wavelength within the wavelength bandwidth to construct the 3D polarization vector field, $\vec{S}(r,\phi,\lambda)$. However, in this work we construct SSVBs whose polarization varies with azimuthal position and along the beam's wavelength spectrum, whilst remaining constant radially outward from the origin. Therefore, it is convenient to omit the radial coordinate and replace it with the spectral coordinate. Effectively, this process involves reducing the initial 3D polarization field $\vec{S}(r,\phi,\lambda)$ to a 2D polarization field $\vec{S}(\lambda,\phi)$ by projecting onto a radial position $r_0$. Since the SSVBs studied in this work do not vary radially, the exact radial position chosen for the projection is done out of convenience. Here we have chosen a radial position where $S_0$ is maximal, thereby ensuring that we are sampling at the point of highest signal-to-noise ratio (SNR). Not only does this allow for a compact representation of the complex tripartite correlations that exist within the beam, it also facilitates numerical computations of the Skyrmion number. \\

\noindent The numerical construction of $\vec{S}(\lambda,\phi)$ is divided into three parts, pictorially shown in three panels in Fig.~\ref{fig:supp_SpatSpecExtract} with the accompanying procedure outlined below:  

\begin{itemize}
    \item[1.] The full spatio-spectral data is organized in the 4D array $\mathbf{S}_{ijkp}$ where the indices $i\in\{1,2,...,N_y\}$ and $j\in\{1,2,...,N_x\}$ correspond to the spatial coordinates, $(x_j,y_i)$, $N_x \! \times \! N_y$ the spatial resolution, $k\in\{1,2,3\}$ the specific Stokes parameter and $p\in\{1,...,N_{\lambda}\}$ the specific wavelength measurement with $N_{\lambda}$ denoting the number of wavelength measurements performed.\\
    
    \item[2.] Next we build the spatio-spectral Stokes vectors, $\mathbf{S}_{ijk}$, element-by-element by projecting along a fixed radial position $r_0$. Each element in this array is constructed by sampling $\mathbf{S}_{ijkp}$ with a fixed annular mask 
    \begin{equation}
     R_{ij}=\begin{cases}1, \; r_0-\frac{\Delta r}{2} < \sqrt{x_j^2 + y_i^2} < r_0+\frac{\Delta r}{2} \\ 0, \; \text{everywhere else}\end{cases}
    \quad,\text{ azimuthal masks}\quad
    \Phi_{ij}^{(l)}=\begin{cases}1, \; \frac{2\pi }{N_\phi}l<\tan^{-1}\left(\frac{y_i}{x_j}\right) <\frac{2\pi}{N_\phi}(l+1)\\ 0, \; \text{everywhere else}\end{cases} .
 \end{equation}
    and at a wavelength $\lambda_p$. Here $\Delta r$ controls the size of the annular mask and $N_{\phi}$ dictates the number of azimuthal segments we want to sample from our field. Here $\Delta r$ can be tuned to optimize for SNR while $N_{\phi}$ should scale to reflect larger variations in the polarization of the field with changing azimuthal angle. In particular, for fields with higher Skyrmion numbers the polarization varies more rapidly in the azimuthal direction thereby necessitating an adequate sampling of the polarization field in the azimuthal direction.
    The resulting coefficient is given by
    \begin{equation}
        A_{kpl} = \frac{1}{M}\sum_{ij} \mathbf{S}_{ijkp} \,R_{ij}\Phi^{(l)}_{ij} =\frac{1}{M}\sum_{ij} \mathbf{S}_{ijkp} \,M^{(l)}_{ij}. 
    \end{equation}
    where $M = \sum_{ij}M_{ij}$. \\ 
    
    \item[3.] Finally the elements of, $\mathbf{S}_{ijk}$, take on the values of $A_{kpl}$ at specific points away from the central index, $(i,j) = \left(\frac{N_y}{2},\frac{N_x}{2}\right)$, corresponding to the chosen spectral measurement, $p$. 
    \begin{equation}
        \mathbf{S}_{ijk} = \sum_{pl} \frac{A_{kpl} \Lambda_{ij}^{(p)} \Phi_{ij}^{(l)}}{\sum_{ij}\Lambda_{ij}^{(p)} \Phi_{ij}^{(l)}}
    \end{equation}
    with $\Lambda_{ij}^{(p)}$ given by
    \begin{equation}
        \Lambda_{ij}^{(p)} = \begin{cases}1, \; c_p-\frac{\Delta c}{2} < \sqrt{x_j^2 + y_i^2} < c_p+\frac{\Delta c}{2} \\ 0, \; \text{everywhere else}\end{cases}
    \end{equation}
    where $c_p$ and $\Delta c$ are tunable parameters controlling the size of the spatio-spectral grid. To match the resolution of the spatial grid the parameters are chosen as $\Delta c=\frac{N_x}{2N_{\lambda}}$ and $c_p=\Delta c \left(p - \frac{1}{2}\right)$. An example of a reconstructed spatio-spectral Stokes parameter, $S_{ij1}$ Stokes parameter is shown in the third panel of Fig~\ref{fig:supp_SpatSpecExtract}. While the azimuthal structure shows 4 azimuthal fringes, the variation is not smooth due to the limited azimuthal sampling of the initial Stokes field. By increasing the number of azimuthal sampling masks, $N_{\phi}$, we can obtain a better approximation to the true continuous spatio-spectral Stokes parameters, an example of which depicted as $S'_{ij1}$ in  Fig~\ref{fig:supp_SpatSpecExtract}.

\begin{figure*}[t!]
\includegraphics[width=\linewidth]{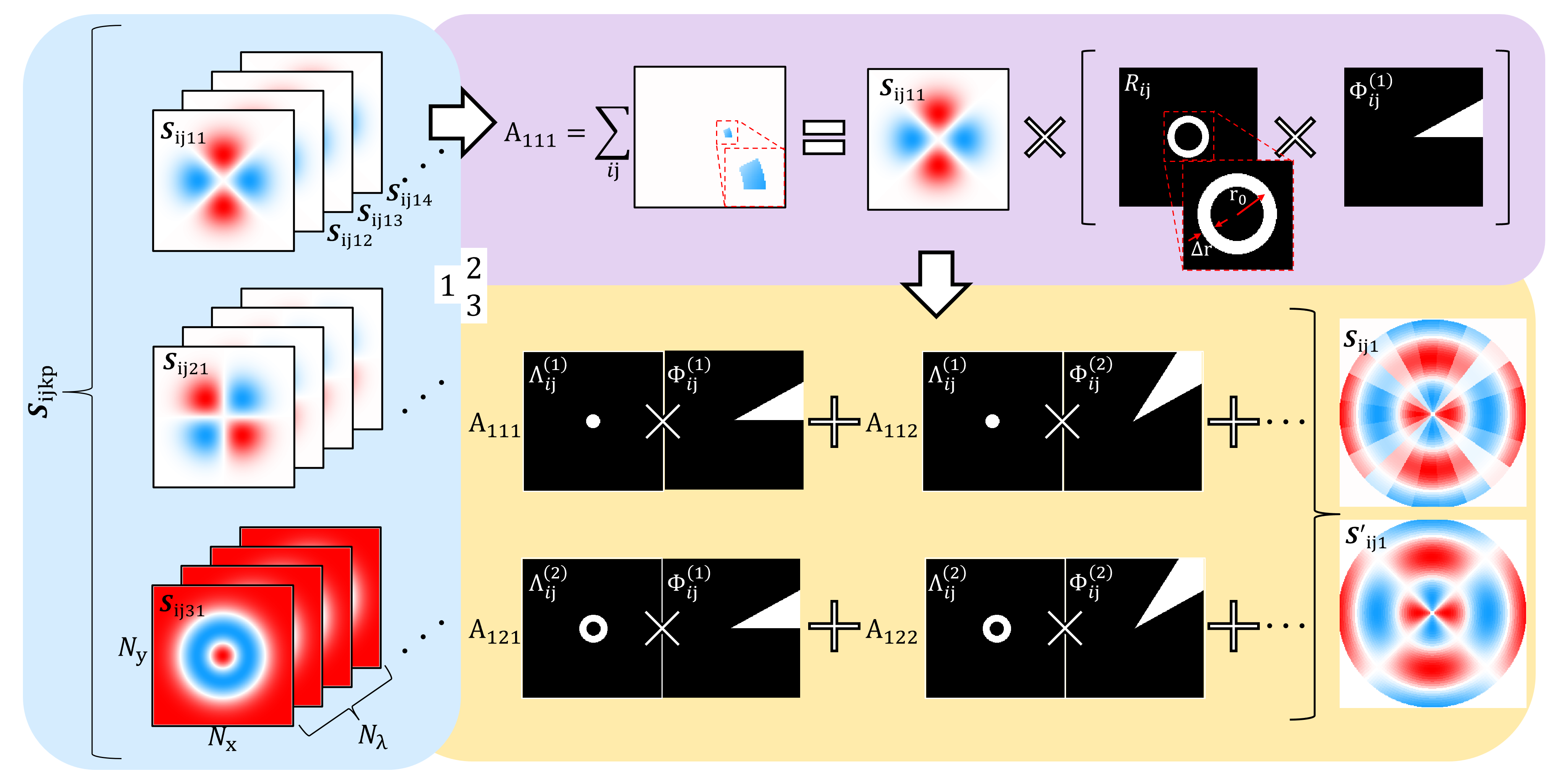}
\caption{\textbf{Extraction of spatio-spectral Stokes parameters}. Panel 1 depicts the organization of the full spatio-spectral Stokes as a 4D array where the first two indices relate to the spatial coordinates, the third coordinate relates to the Stokes parameter and the fourth index relates to the spectral coordinate. In panel 2 the extraction of the coeffecients, $A_{kpl}$, are shown with the example shown for the coeffecient $A_{111}$. The coeffecients are calculated by averaging over local regions of $S_{ijkp}$ selected by multiplying it by annular masks, $R_{ij}$, and azimuthal masks $\Phi_{ij}^{(l)}$, with zoomed-in images of the masks shown as insets. Panel 3 shows how each spatio-spectral parameter, $S_{ijk}$, is calculated through the multiplication of the masks $\Lambda_{ij}^p\Phi_{ij}^l$ with the previously calculated coeffecients.}
\label{fig:supp_SpatSpecExtract}
\end{figure*}

\end{itemize}


 



\end{widetext}


\newpage
\begin{thebibliography}{10}

\bibitem{skyrme1962unified}
T.~H.~R. Skyrme, ``A unified field theory of mesons and baryons,'' {\em Nuclear Physics}, vol.~31, pp.~556--569, 1962.

\bibitem{ZAHED19861}
I.~Zahed and G.~Brown, ``The skyrme model,'' {\em Physics Reports}, vol.~142, no.~1, pp.~1--102, 1986.

\bibitem{Naya2018Skyrmions}
C.~Naya and P.~Sutcliffe, ``Skyrmions and clustering in light nuclei,'' {\em Physical review letters}, vol.~121, no.~23, p.~232002, 2018.

\bibitem{eisenberg1981nucleon}
J.~Eisenberg and G.~Kälbermann, ``The use of skyrmions for two-nucleon systems,'' {\em Progress in Particle and Nuclear Physics}, vol.~22, pp.~1--42, 1989.

\bibitem{leslie2009creation}
L.~Leslie, A.~Hansen, K.~Wright, B.~Deutsch, and N.~Bigelow, ``Creation and detection of skyrmions in a bose-einstein condensate,'' {\em Physical review letters}, vol.~103, no.~25, p.~250401, 2009.

\bibitem{ackerman2015self}
P.~J. Ackerman, J.~Van De~Lagemaat, and I.~I. Smalyukh, ``Self-assembly and electrostriction of arrays and chains of hopfion particles in chiral liquid crystals,'' {\em Nature communications}, vol.~6, no.~1, pp.~1--9, 2015.

\bibitem{ge2021observation}
H.~Ge, X.-Y. Xu, L.~Liu, R.~Xu, Z.-K. Lin, S.-Y. Yu, M.~Bao, J.-H. Jiang, M.-H. Lu, and Y.-F. Chen, ``Observation of acoustic skyrmions,'' {\em Physical Review Letters}, vol.~127, no.~14, p.~144502, 2021.

\bibitem{muelas2022observation}
R.~D. Muelas-Hurtado, K.~Volke-Sep{\'u}lveda, J.~L. Ealo, F.~Nori, M.~A. Alonso, K.~Y. Bliokh, and E.~Brasselet, ``Observation of polarization singularities and topological textures in sound waves,'' {\em Physical Review Letters}, vol.~129, no.~20, p.~204301, 2022.

\bibitem{wang2025topological}
B.~Wang, Z.~Che, C.~Cheng, C.~Tong, L.~Shi, Y.~Shen, K.~Y. Bliokh, and J.~Zi, ``Topological water-wave structures manipulating particles,'' {\em Nature}, pp.~1--7, 2025.

\bibitem{tsesses2018optical}
S.~Tsesses, E.~Ostrovsky, K.~Cohen, B.~Gjonaj, N.~Lindner, and G.~Bartal, ``Optical skyrmion lattice in evanescent electromagnetic fields,'' {\em Science}, vol.~361, no.~6406, pp.~993--996, 2018.

\bibitem{lei2025skyrmionic}
X.~Lei, A.~Yang, X.~Chen, L.~Du, P.~Shi, Q.~Zhan, and X.~Yuan, ``Skyrmionic spin textures in nonparaxial light,'' {\em Advanced Photonics}, vol.~7, no.~1, pp.~016009--016009, 2025.

\bibitem{du2019deep}
L.~Du, A.~Yang, A.~V. Zayats, and X.~Yuan, ``Deep-subwavelength features of photonic skyrmions in a confined electromagnetic field with orbital angular momentum,'' {\em Nature Physics}, vol.~15, no.~7, pp.~650--654, 2019.

\bibitem{krol2021observation}
M.~Kr{\'o}l, H.~Sigurdsson, K.~Rechci{\'n}ska, P.~Oliwa, K.~Tyszka, W.~Bardyszewski, A.~Opala, M.~Matuszewski, P.~Morawiak, R.~Mazur, {\em et~al.}, ``Observation of second-order meron polarization textures in optical microcavities,'' {\em Optica}, vol.~8, no.~2, pp.~255--261, 2021.

\bibitem{shen2021supertoroidal}
Y.~Shen, Y.~Hou, N.~Papasimakis, and N.~I. Zheludev, ``Supertoroidal light pulses as electromagnetic skyrmions propagating in free space,'' {\em Nature communications}, vol.~12, no.~1, p.~5891, 2021.

\bibitem{cisowski2023building}
C.~Cisowski, C.~Ross, and S.~Franke-Arnold, ``Building paraxial optical skyrmions using rational maps,'' {\em Advanced Photonics Research}, vol.~4, no.~4, p.~2200350, 2023.

\bibitem{shen2022generation}
Y.~Shen, E.~C. Mart{\'\i}nez, and C.~Rosales-Guzm{\'a}n, ``Generation of optical skyrmions with tunable topological textures,'' {\em Acs Photonics}, vol.~9, no.~1, pp.~296--303, 2022.

\bibitem{gao2020paraxial}
S.~Gao, F.~C. Speirits, F.~Castellucci, S.~Franke-Arnold, S.~M. Barnett, and J.~B. G{\"o}tte, ``Paraxial skyrmionic beams,'' {\em Physical Review A}, vol.~102, no.~5, p.~053513, 2020.

\bibitem{singh2023synthetic}
K.~Singh, P.~Ornelas, A.~Dudley, and A.~Forbes, ``Synthetic spin dynamics with bessel-gaussian optical skyrmions,'' {\em Optics Express}, vol.~31, no.~10, pp.~15289--15300, 2023.

\bibitem{teng2023physical}
H.~Teng, J.~Zhong, J.~Chen, X.~Lei, and Q.~Zhan, ``Physical conversion and superposition of optical skyrmion topologies,'' {\em Photonics Research}, vol.~11, no.~12, pp.~2042--2053, 2023.

\bibitem{shen2021topological}
Y.~Shen, ``Topological bimeronic beams,'' {\em Optics Letters}, vol.~46, no.~15, pp.~3737--3740, 2021.

\bibitem{ma2025nanophotonic}
J.~Ma, J.~Yang, S.~Liu, B.~Chen, X.~Li, C.~Song, G.~Qiu, K.~Zou, X.~Hu, F.~Li, {\em et~al.}, ``Nanophotonic quantum skyrmions enabled by semiconductor cavity quantum electrodynamics,'' {\em Nature Physics}, vol.~21, no.~9, pp.~1462--1468, 2025.

\bibitem{koni2025dual}
M.~Koni, F.~Nothlawala, V.~Hakobyan, I.~Nape, E.~Brasselet, and A.~Forbes, ``Dual-wavelength quantum skyrmions from liquid crystal topological defects,'' {\em Physical Review Letters}, vol.~135, no.~22, p.~223804, 2025.

\bibitem{ornelas2024non}
P.~Ornelas, I.~Nape, R.~de~Mello~Koch, and A.~Forbes, ``Non-local skyrmions as topologically resilient quantum entangled states of light,'' {\em Nature Photonics}, vol.~18, no.~3, pp.~258--266, 2024.

\bibitem{wang2025generation}
J.~Wang, X.~Zeng, K.~Ren, Z.~Ye, C.~M. Cisowski, Y.~Chen, X.~Yang, C.~Wang, H.~Gao, and S.~Franke-Arnold, ``Generation of ring-shaped optical skyrmion with a high topological number,'' {\em Applied Physics Letters}, vol.~126, no.~20, 2025.

\bibitem{zeng2025tailoring}
X.~Zeng, J.~Fang, H.~Wu, J.~Wang, Y.~Chen, Y.~Zhou, X.~Yang, C.~Wang, D.~Wei, H.~Chen, {\em et~al.}, ``Tailoring ultra-high-order optical skyrmions,'' {\em Laser \& Photonics Reviews}, vol.~19, no.~21, p.~e00732, 2025.

\bibitem{wang2025perturbation}
A.~A. Wang, Y.~Ma, Y.~Zhang, Z.~Zhao, Y.~Cai, X.~Qiu, B.~Dong, and C.~He, ``Perturbation-resilient integer arithmetic using optical skyrmions,'' {\em Nature Photonics}, pp.~1--9, 2025.

\bibitem{wang2024topological}
A.~A. Wang, Z.~Zhao, Y.~Ma, Y.~Cai, R.~Zhang, X.~Shang, Y.~Zhang, J.~Qin, Z.-K. Pong, T.~Marozs{\'a}k, {\em et~al.}, ``Topological protection of optical skyrmions through complex media,'' {\em Light: Science \& Applications}, vol.~13, no.~1, p.~314, 2024.

\bibitem{ornelas2025topological}
P.~Ornelas, I.~Nape, R.~de~Mello~Koch, and A.~Forbes, ``Topological rejection of noise by quantum skyrmions,'' {\em Nature Communications}, vol.~16, no.~1, p.~2934, 2025.

\bibitem{de2025quantum}
R.~de~Mello~Koch, B.-Q. Lu, P.~Ornelas, I.~Nape, and A.~Forbes, ``Quantum skyrmions in general quantum channels,'' {\em APL Quantum}, vol.~2, no.~2, 2025.

\bibitem{guo2026topological}
Z.~Guo, C.~Peters, N.~Mata-Cervera, A.~N. Vetlugin, R.~Guo, P.~Zhang, A.~Forbes, and Y.~Shen, ``Topological robustness of classical and quantum optical skyrmions in atmospheric turbulence,'' {\em Nature Communications}, 2026.

\bibitem{sugic2021particle}
D.~Sugic, R.~Droop, E.~Otte, D.~Ehrmanntraut, F.~Nori, J.~Ruostekoski, C.~Denz, and M.~R. Dennis, ``Particle-like topologies in light,'' {\em Nature communications}, vol.~12, no.~1, pp.~1--10, 2021.

\bibitem{yao2024multi}
J.~Yao, Y.~Shen, J.~Hu, and Y.~Yang, ``Multi-degree-of-freedom hybrid optical skyrmions,'' {\em arXiv preprint arXiv:2409.05689}, 2024.

\bibitem{chong2020generation}
A.~Chong, C.~Wan, J.~Chen, and Q.~Zhan, ``Generation of spatiotemporal optical vortices with controllable transverse orbital angular momentum,'' {\em Nature Photonics}, vol.~14, no.~6, pp.~350--354, 2020.

\bibitem{teng2025construction}
H.~Teng, X.~Liu, N.~Zhang, H.~Fan, G.~Chen, Q.~Cao, J.~Zhong, X.~Lei, and Q.~Zhan, ``Construction of optical spatiotemporal skyrmions,'' {\em Light: Science \& Applications}, vol.~14, no.~1, p.~324, 2025.

\bibitem{kopf2023correlating}
L.~Kopf, R.~Barros, and R.~Fickler, ``Correlating space, wavelength, and polarization of light: Spatiospectral vector beams,'' {\em ACS Photonics}, vol.~11, no.~1, pp.~241--246, 2023.

\bibitem{fickler2024higher}
R.~Fickler, L.~Kopf, and M.~Ornigotti, ``Higher-order poincar{\'e} spheres and spatiospectral poincar{\'e} beams,'' {\em Physical Review Research}, vol.~6, no.~3, p.~033298, 2024.

\bibitem{shen2023optical}
Y.~Shen, Q.~Zhang, P.~Shi, L.~Du, X.~Yuan, and A.~V. Zayats, ``Optical skyrmions and other topological quasiparticles of light,'' {\em Nature Photonics}, pp.~1--11, 2023.

\bibitem{zhan2024spatiotemporal}
Q.~Zhan, ``Spatiotemporal sculpturing of light: a tutorial,'' {\em Advances in Optics and Photonics}, vol.~16, no.~2, pp.~163--228, 2024.

\bibitem{shen2024optical}
Y.~Shen, Q.~Zhang, P.~Shi, L.~Du, X.~Yuan, and A.~V. Zayats, ``Optical skyrmions and other topological quasiparticles of light,'' {\em Nature Photonics}, vol.~18, no.~1, pp.~15--25, 2024.

\bibitem{dennis2009singular}
M.~R. Dennis, K.~O'holleran, and M.~J. Padgett, ``Singular optics: optical vortices and polarization singularities,'' in {\em Progress in optics}, vol.~53, pp.~293--363, Elsevier, 2009.

\bibitem{cardano2013generation}
F.~Cardano, E.~Karimi, L.~Marrucci, C.~De~Lisio, and E.~Santamato, ``Generation and dynamics of optical beams with polarization singularities,'' {\em Optics express}, vol.~21, no.~7, pp.~8815--8820, 2013.

\bibitem{otte2018polarization}
E.~Otte, C.~Alpmann, and C.~Denz, ``Polarization singularity explosions in tailored light fields,'' {\em Laser \& Photonics Reviews}, vol.~12, no.~6, p.~1700200, 2018.

\bibitem{larocque2018reconstructing}
H.~Larocque, D.~Sugic, D.~Mortimer, A.~J. Taylor, R.~Fickler, R.~W. Boyd, M.~R. Dennis, and E.~Karimi, ``Reconstructing the topology of optical polarization knots,'' {\em Nature Physics}, vol.~14, no.~11, pp.~1079--1082, 2018.

\bibitem{shen2023topological}
Y.~Shen, B.~Yu, H.~Wu, C.~Li, Z.~Zhu, and A.~V. Zayats, ``Topological transformation and free-space transport of photonic hopfions,'' {\em Advanced Photonics}, vol.~5, no.~1, p.~015001, 2023.

\bibitem{wan2022scalar}
C.~Wan, Y.~Shen, A.~Chong, and Q.~Zhan, ``Scalar optical hopfions,'' {\em Elight}, vol.~2, no.~1, p.~22, 2022.

\bibitem{ehrmanntraut2023optical}
D.~Ehrmanntraut, R.~Droop, D.~Sugic, E.~Otte, M.~R. Dennis, and C.~Denz, ``Optical second-order skyrmionic hopfion,'' {\em Optica}, vol.~10, no.~6, pp.~725--731, 2023.

\end{thebibliography}

\begin{thebibliography}{1}

\bibitem{tamovsauskas2018transmittance}
G.~Tamo{\v{s}}auskas, G.~Beresnevi{\v{c}}ius, D.~Gadonas, and A.~Dubietis, ``Transmittance and phase matching of bbo crystal in the 3- 5 $\mu$ m range and its application for the characterization of mid-infrared laser pulses,'' {\em Optical Materials Express}, vol.~8, no.~6, pp.~1410--1418, 2018.

\bibitem{sakakura2020ultralow}
M.~Sakakura, Y.~Lei, L.~Wang, Y.-H. Yu, and P.~G. Kazansky, ``Ultralow-loss geometric phase and polarization shaping by ultrafast laser writing in silica glass,'' {\em Light: Science \& Applications}, vol.~9, no.~1, p.~15, 2020.

\bibitem{della2008micromachining}
G.~Della~Valle, R.~Osellame, and P.~Laporta, ``Micromachining of photonic devices by femtosecond laser pulses,'' {\em Journal of Optics A: Pure and Applied Optics}, vol.~11, no.~1, p.~013001, 2008.

\bibitem{sugioka2014ultrafast}
K.~Sugioka and Y.~Cheng, ``Ultrafast lasers—reliable tools for advanced materials processing,'' {\em Light: Science \& Applications}, vol.~3, no.~4, pp.~e149--e149, 2014.

\end{thebibliography}
\end{document}